%% file: main.tex
\documentclass[a4paper,11pt]{article}
\pdfoutput=1
\usepackage{jcappub}
\input{preamble}

\usepackage[utf8x]{inputenc}
\usepackage[normalem]{ulem}

\begin{document}

\title{Cosmic-ray propagation with DRAGON2: II.~Nuclear interactions with the interstellar gas.}

\author[a]{Carmelo Evoli}
\emailAdd{carmelo.evoli@gssi.it}
\note{C.~Evoli:~\href{http://orcid.org/0000-0002-6023-5253/}{orcid.org/0000-0002-6023-5250}}  
\author[b]{Daniele Gaggero}
\emailAdd{d.gaggero@uva.nl}
\author[c,d]{Andrea Vittino}
\emailAdd{vittino@physik.rwth-aachen.de}
\author[e]{Mattia Di Mauro}
\emailAdd{mdimauro@slac.stanford.edu}
\author[f]{Dario Grasso}
\emailAdd{dario.grasso@pi.infn.it}
\author[g]{Mario Nicola Mazziotta}
\emailAdd{mazziotta@ba.infn.it}

\affiliation[a]{\small Gran Sasso Science Institute, Viale Francesco Crispi 7, 67100 L'Aquila, Italy \\ INFN/Laboratori Nazionali del Gran Sasso, Via G. Acitelli 22, 67100, Assergi (AQ), Italy}
\affiliation[b]{\small GRAPPA, Institute of Physics, University of Amsterdam \\ 1098 XH Amsterdam, The 
Netherlands}
\affiliation[c]{\small Physik-Department T30d, Technische Universit{\"a}t M{\"u}nchen, James Franck-Str. 1, D-85748, Garching, Germany}
\affiliation[d]{\small Institute for Theoretical Particle Physics and Cosmology (TTK), RWTH Aachen University, D-52056 Aachen, Germany}
\affiliation[e]{\small W.~W.~Hansen Experimental Physics Laboratory, Kavli Institute for Particle Astrophysics and Cosmology, Department of Physics and SLAC National Accelerator Laboratory \\
Stanford University, Stanford, CA 94305, USA}
\affiliation[f]{\small INFN Pisa, Largo B. Pontecorvo 3, I-56127 Pisa, Italy}
\affiliation[g]{\small Istituto Nazionale di Fisica Nucleare, Sezione di Bari, 70126 Bari, Italy}

\date{\today}


\arxivnumber{TUM-HEP 1111/17, TTK-17-40}

\begin{abstract}
{Understanding the isotopic composition of cosmic rays (CRs) observed near Earth represents a milestone towards the identification of their origin.
Local fluxes contain all the known stable and long-lived isotopes, reflecting the complex history of primaries and secondaries as they traverse the interstellar medium.
For that reason, a numerical code which aims at describing the CR transport in the Galaxy must unavoidably rely on accurate modelling of the production of secondary particles.

In this work we provide a detailed description of the nuclear cross sections and decay network as implemented in the forthcoming release of the galactic propagation code~\dragon.

We present the secondary production models implemented in the code and we apply the different prescriptions to compute quantities of interest to interpret local CR fluxes~(e.g., nuclear fragmentation timescales, secondary and tertiary source terms).
In particular, we develop a nuclear secondary production model aimed at accurately computing the light secondary fluxes (namely: Li, Be, B) above 1 GeV/n.
This result is achieved by fitting existing empirical or semi-empirical formalisms to a large sample of measurements in the energy range 100 MeV/n to 100 GeV/n and by considering the contribution of the most relevant decaying isotopes up to iron.
Concerning secondary antiparticles (positrons and antiprotons), we describe a collection of models taken from the literature, and provide a detailed quantitative comparison.}
\end{abstract}

\maketitle

\newpage

\input{introduction}
\input{secondary}
\input{leptons}
\input{antiprotons}
\input{conclusions}

\input{acknowledgments}

\bibliographystyle{JHEP}
\bibliography{dragon_paper_II}

\end{document}

%% file: preamble.tex
\usepackage{amsmath}
\usepackage{amssymb}
\usepackage{amsfonts}
\usepackage{mathtools}

\usepackage{lineno}
\usepackage{lscape}
\usepackage{soul}
\usepackage{subfigure}
\usepackage{graphicx}
\usepackage{color}
\usepackage{bm}
\usepackage{epstopdf}
\usepackage{url}
\usepackage{booktabs}
\usepackage{mdframed}
\usepackage{hyperref}
\usepackage{booktabs}
\usepackage{siunitx}
\usepackage{multirow}

\newcommand*{\dragon}{{\tt DRAGON2}}

\newcommand*{\galprop}{{\tt GALPROP}}
\newcommand*{\exfor}{{\tt EXFOR}}

%% file: introduction.tex
\section{Introduction}\label{sec:introduction}

The latest years have witnessed an extraordinary upsurge in the quantity and accuracy of cosmic-ray (CR) observations. 
The dramatic increase in the quality of data, when considering the impressive experimental efforts from the present and next generation of instruments (e.g., AMS-02~\cite{2013PhRvL.110n1102A}, BESS~\cite{2017AdSpR..60..806A}, CALET~\cite{CALET}, DAMPE~\cite{2017APh....95....6C}, ISS-CREAM~\cite{2014AdSpR..53.1451S}, PAMELA~\cite{2013PhRvL.111h1102A}), has emphasised the necessity of a much more accurate and realistic modeling of CR transport in the Galaxy. 
With this ambitious goal in mind, we are currently working on the development of~{\tt DRAGON2}, the new version of the public code~{\tt DRAGON}.

{\tt DRAGON2} serves as a tool to solve the CR transport equation for all CR species. The project mainly focuses on CRs produced in astrophysical sources, such as supernova remnants (SNRs) or pulsars, and also accounts for an hypothetical contribution from beyond-standard-model processes (e.g., particle dark matter annihilation/decay). In this effort {\tt DRAGON2} joins a list of numerical packages that originated from the well-known {\tt GALPROP}\footnote{Publicly available from \url{galprop.stanford.edu}. An extended version of GALPROPv54 can be found at~\url{gitlab.mpcdf.mpg.de/aws/galprop}~\cite{2015ICRC...34..507S}} code~\cite{Galprop1,Galprop2,Galprop3,2001ICRC....5.1942S}, and that includes the {\tt USINE}~\cite{Usine} and {\tt PICARD}~\cite{Picard1,Picard2} packages. Among the peculiar features of {\tt DRAGON2}, we remark the  modeling of the coefficients appearing in the transport equation, which are implemented as fully position- and energy-dependent operators and are discretized on a non-uniform and adaptable spatial grid. This makes {\tt DRAGON2} a suitable tool to model CR transport in a three-dimensional and anisotropic setup. Moreover, {\tt DRAGON2} adopts a time-integrating scheme for the transport equation and can therefore operate also beyond the steady-state assumption, e.g., to model CR transients (see also~\cite{2001ICRC....5.1964S} for time-dependent solutions obtained with~\galprop).

This work is the second in a series of papers aimed at presenting {\tt DRAGON2}. 

While in~\cite{2017JCAP...02..015E} we provided a description of the transport equation solver, and discussed all the relevant astrophysical ingredients related to the modeling of CR injection and energy losses, here we focus our attention on the topic of secondary CRs, i.e., the particles and antiparticles that are produced as the result of a complex network of radioactive decays and spallation reactions taking place during the journey of primary CRs across the Interstellar Medium (ISM) of our Galaxy. Due to this composite chain of processes, the spectrum of a given isotope includes contributions from the primary CR source and from the fragmentation of many different parents, that can in turn fragment into the observed isotope: This pattern of multiple parentage gives to each observed isotope a potentially very complex history (see~\cite{1984ApJS...56..369L} as a seminal work for modelling propagation of heavier than helium nuclei).
 
On the other hand, secondary antiparticles -- antiprotons and positrons being the prime example -- enter as an irreducible background in the search for imprints of new physics (e.g., dark matter annihilation/decay~\cite{2015JCAP...09..023G,Evoli:2015vaa,Winkler:2017xor}) or in the study of non-standard astrophysical processes (e.g., the hadronic production and acceleration in aged SNRs~\cite{2009PhRvL.103e1104B}).

In both situations, an accurate knowledge of nuclear cross sections is required.

It is with this necessity in mind that we have approached the modeling of the spallation and decay processes in {\tt DRAGON2}. 

In fact, while the previous version of our code benefited from using the cross section routines and the nuclear reaction network derived from the \galprop~code\footnote{With the exception of~\cite{2016APh....81...21M} where a comprehensive calculation of the secondary yields were calculated using FLUKA.}, within the {\tt DRAGON2} project we have performed a critical re-evaluation of the main models in the literature, and developed an independent algorithm to compute the relevant nuclear cross sections. 
This paper contains then a detailed and fully reproducible description of the cross-section network as implemented in our code, and presents a comprehensive comparison with an updated database of experimental measurements.

In an effort to meet the needs of the community of people working in the field of CR physics, and as an anticipation of the upcoming {\tt DRAGON2} release, this paper is accompanied by the numerical library {\tt D2XSECS}\footnote{It can be downloaded from  \url{https://github.com/cosmicrays/D2XSECS}}, and which can be used to compute all the cross sections discussed here. 

\section{The nuclear chain}

For a CR species $i$ with atomic number $Z_i$, mass number $A_i$, kinetic energy per nucleon $T$ and total momentum $p$, the number density per total momentum $N_i(T, \vec{x})$ in the Galaxy is described by the steady-state solution of the transport equation~\cite{Berezinskii1990}:
\begin{equation}
\begin{split}
{\bf \nabla} \cdot (\vec{J}_i - {\vec{v}_w} N_i) + \frac{\partial}{\partial p} \left[ p^2 D_{pp} \, \frac{\partial}{\partial p} \left( \frac{N_i}{p^2} \right) \right] - \frac{\partial}{\partial p} \left[ \dot{p} N_i - \frac{p}{3} \left(\vec{\nabla} \cdot \vec{v}_w \right) N_i \right] = \\
Q_{\rm source} 
- \frac{N_i}{\tau^{\rm f}_i} 
+ \sum_{j} \Gamma^{\rm s}_{j \rightarrow i}(N_j)
- \frac{N_i}{\tau^{\rm r}_i} 
+ \sum_{j} \frac{N_j}{\tau^{\rm r}_{j \rightarrow i}} 
\end{split}\label{eq:propeq}
\end{equation}
where $\vec{J}$ is the diffusive flux, $D_{pp}$ is the momentum diffusion coefficient, $Q_{\rm source}$ describes the distribution and the energy spectra of primary sources, $\vec{v}_w (\vec{r})$ is the Galactic wind velocity responsible for CR advection and $\dot{p}(\vec{r},p)$ accounts for the momentum losses. See~\cite{2017JCAP...02..015E} for more details.

The fragmentation timescale, $\tau^{\rm f}_i$, is associated with the total inelastic scattering of a nucleus $i$ with the interstellar gas targets, while $\Gamma^{\rm s}_{j \rightarrow i}$ describes the source term of a secondary nucleus $i$ by spallation of a heavier species $j$. 
The summations in equation~(\ref{eq:propeq}) are over all CR species heavier than $i$. 

In their most general form, $\tau^{\rm f}_i$ and $\Gamma^{\rm s}_{j \rightarrow i}$ can be defined as:
\begin{equation}\label{eq:fragmentation}
\frac{1}{\tau^{\rm f}_i(T)} = \beta(T) c n_{\rm H} \left[ \sigma_{\rm H,i}(T) + f_{\rm He} \, \sigma_{\rm He,i}(T) \right]
\end{equation}
and
\begin{equation}\label{eq:secondary}
\Gamma^{\rm s}_{j \rightarrow i}(T) = c n_{\rm H} \int \, dT' \beta(T') N_j(T') \left[ \frac{d\sigma_{\rm H,j \rightarrow i}}{dT}(T,T') + f_{\rm He} \frac{d\sigma_{\rm He,j \rightarrow i}}{dT}(T, T')  \right]
\end{equation}
where $T'$ is the kinetic energy per nucleon of the parent particle, $n_{\rm H} = n_{\rm HI} + 2 n_{\rm H2} + n_{\rm HII}$ is the interstellar hydrogen density and $f_{\rm He} \equiv n_{\rm He} / n_{\rm H} = 0.11$ is the helium fraction (by number). In doing so, we assume constant H-to-He ratio in the ISM and we ignore the contribution of heavier elements as targets.

The inelastic cross sections $\sigma_{\rm k,i}(T)$ in equation~(\ref{eq:fragmentation}) represent the probability that the CR nucleus $i$ undergoes a reaction with an interstellar nucleus of kind $k$ (see section \ref{sec:inelastic}). 
Since in this interaction the incoming and outgoing particles differ in most cases, this process is an effective ``sink'' for the CR species $i$.
Equation~(\ref{eq:secondary}) represents the opposite process, that is the gain of CRs of kind $i$ as a result of the fragmentation of the heavier species $j$.
It involves the partial interaction cross sections
$\frac{d\sigma_{\rm k,j \rightarrow i}}{dT}$ which describe the production of $i$ in processes where CRs  of the species $j$ collide with the hydrogen and helium of the ISM.   

Lastly, $\tau^{\rm r}_i$ is the lifetime of a nucleus of type $i$ with respect to radioactive decays (and it is infinite for stable nuclei), while the last term in equation~(\ref{eq:propeq}) describes the appearance of nuclei $i$ due to decays of other nuclei.
The typical disintegration modes are: $\beta^\pm$ decay, where the nucleus decays spontaneously, and electron-capture (EC) decay where the nucleus decays by capturing an electron after one has been previously attached~\cite{1985Ap&SS.114..365L}. 
In the latter case, the decay rate depends on the ambient electron density and on the typical attachment time. 

Since the lighter species can originate from the spallation or decay of the heavier ones, we start the evaluation of CR number densities from the heaviest primary.
Then, we compute the flux of the next-to-heaviest nucleus, whose spallation term only depends on the heaviest (already evaluated the previous step). 
This procedure is iterated for all the nuclei in reverse mass order, all the way down to Hydrogen ($A=1$).
To account for beta-decays, we repeat twice the propagation of beta-coupled nuclei as described in section~\ref{sec:decays}.
As a last step, electrons, positrons (see section~\ref{sec:leptons}) and antiprotons (see section~\ref{sec:antiprotons}) are propagated. 

We start the nuclear network with Iron ($Z=26$), since heavier species contribute to the LiBeB source term for less than 1\%.

%% file: secondary.tex
\section{Inelastic scattering}\label{sec:inelastic}

The inelastic cross section as a function of energy entering equation~(\ref{eq:fragmentation}) is usually given in terms of semi-empirical formul\ae,  which typically reproduce all existing data quite well, while first-principle approaches cannot be considered fully successful so far~(see the discussion in~\cite{2002astro.ph.12111M}). 

\begin{figure}
\begin{center}
\includegraphics[width=0.48\columnwidth]{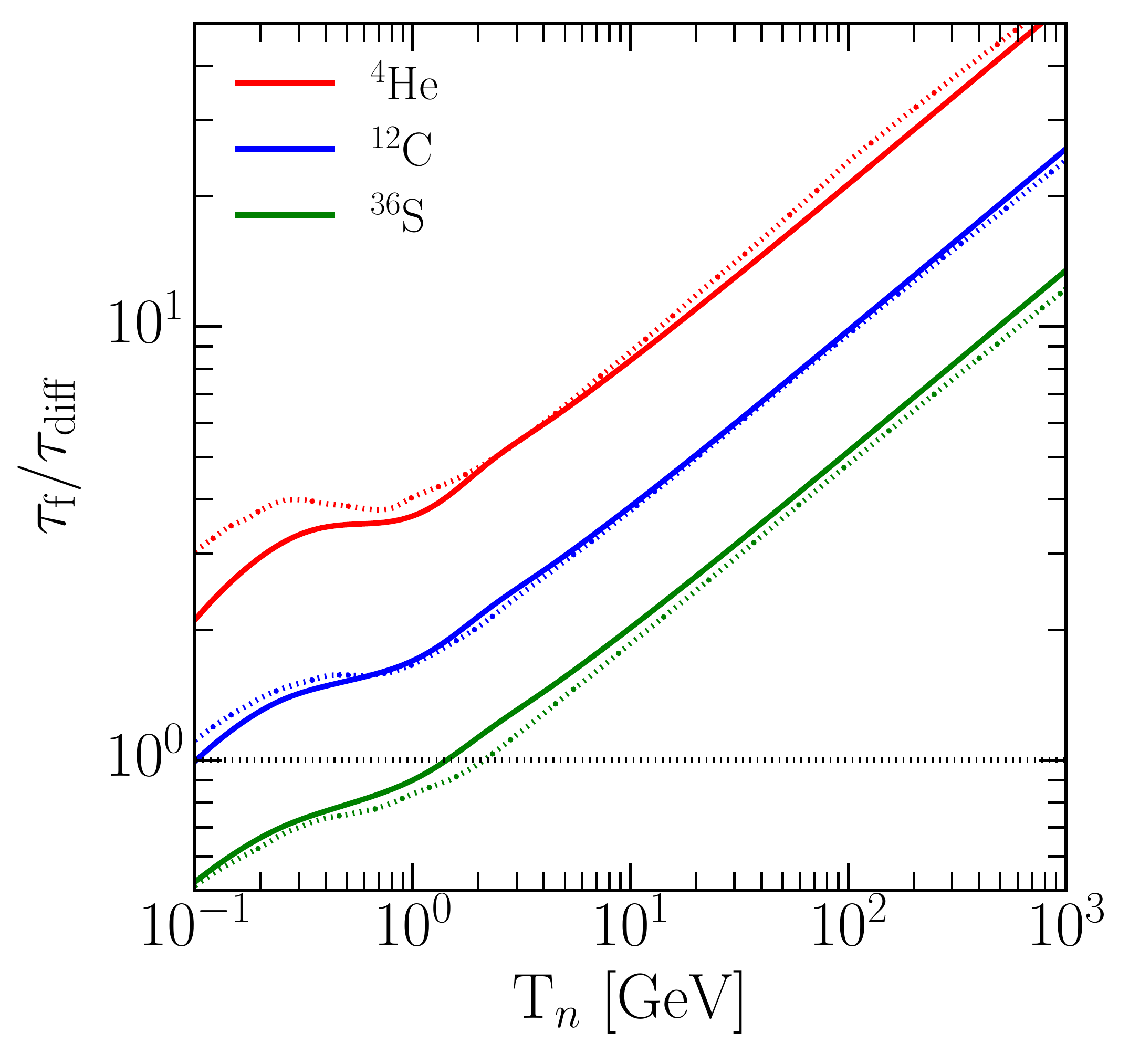} 
\end{center}
\caption{Ratio between the fragmentation and diffusion timescales for the Tripathi99 (solid lines) and the CROSEC (dotted lines) models.
The diffusion model parameters used here are: $D_0/H = 0.75 \times 10^{28}$~cm$^2$/s/kpc, $\delta = 0.42$, $v_A = 27$~km/s, $dV_C/dz = 14$~km/s/kpc~\cite{Evoli:2015vaa}.}
\label{fig:inelastic_all}
\end{figure}

We describe below two different parameterizations implemented in the \dragon~code. 
In figure~\ref{fig:inelastic_all} we show the ratio between the fragmentation timescale (computed with the two different models) and the diffusive escape timescale (obtained from fitting recent AMS-02 data in~\cite{Evoli:2015vaa}\footnote{The transport parameters corresponding to the best fit are: $D_0/H = 0.75 \times 10^{28}$~cm$^2$/s/kpc, $\delta = 0.42$, $v_A = 27$~km/s, $dV_C/dz = 14$~km/s/kpc.}). % 
The ratio is computed under the assumption that the gas (where fragmentation takes place) uniformly fills a disk of height $h = 100$~pc with density $n_{\rm H} = 1$~cm$^{-3}$.
Moreover, to compute the relative contribution of He in the target, we adopt the phenomenological scaling proposed in~\cite{1988PhRvC..37.1490F}:
\begin{equation}
\frac{\sigma_{{\rm He},i}}{\sigma_{{\rm H},i}} = 2.10 \, A_i^{0.055}
\end{equation}

As it can be seen in figure~\ref{fig:inelastic_all}, the heavier nuclei have larger cross sections and break more easily during their wandering across the Galaxy.
As a consequence, fragmentation is negligible  with respect to diffusion for lighter nuclei ($Z \lesssim 6$), becoming more and more important for heavier species. 
At high energies ($T > 2$~GeV/n) the two parametrizations are nearly constant and in good agreement with each other, while at lower energies they may differ up to $\sim 20$\%. 

\begin{description}

\item[Tripathi99] {We describe here our implementation of this model, based on ~\cite{Tripathi1997a,Tripathi1997b,Tripathi1999}} 

The formulation is similar to the geometrical picture in which the target ($t$) and projectile ($p$) areas are summed with some corrections:
\begin{equation}\label{eq:tripathi}
\sigma_{t,p} = \pi r_0^2 (A_p^{1/3} + A_t^{1/3} + \delta_E)^2 \left( 1 - \frac{R_c B}{E^{\rm cm}_{\rm MeV}} \right) % X_m
\end{equation}
where $r_0 = 1.1$~fm and $E^{\rm cm}$ is the center of mass energy of the colliding system . 

The correction term $\delta_E$ takes into account the overlapping volume of the colliding system, the energy dependence of nucleus transparency and isospin effects:
\begin{equation}
\delta_E = 1.85 S + 0.16 S \left( E^{\rm cm}_{\rm MeV}\right)^{-1/3} - C_E + 0.91 \frac{(A_t - 2 Z_t)Z_p}{A_p A_t}
\end{equation}
where $S$ is the mass asymmetry term and is given by:
\begin{equation}
S = \frac{A_t^{1/3}A_p^{1/3}}{(A_t^{1/3}+A_p^{1/3})}
\end{equation}
The term $C_E$ is related to the transparency and Pauli blocking and is given by:
\begin{equation}
C_E = D \left[ 1 - \exp \left(-\frac{T}{T_1} \right)\right] 
- 0.292 \exp \left( - \frac{T}{792 \, {\rm MeV/n}} \right) 
\cos \left[ 0.229 \, \left(\frac{T}{\rm MeV/n} \right)^{0.453} \right]
\end{equation}
where $T_1$ and $D$ are parameters related to the specific reaction and are given in ~\cite{Tripathi1999}.

In equation~(\ref{eq:tripathi}), $B$ is the energy-dependent Coulomb interaction barrier:
\begin{equation}
B = \frac{1.44 Z_t Z_p}{R}
\end{equation}
where $R$ is the distance at which the barrier is evaluated, 

\begin{equation}
R = r_p + r_t + 1.2(A_t^{1/3}+A_p^{1/3}) \left(E^{\rm cm}_{\rm MeV} \right)^{-1/3}
\end{equation}
with $r_{p}$($r_{t}$) being the equivalent radius of the hard sphere of projectile (target) nucleus for which we take $r(A) = 1.29$~fm $A^{1/3}$.

Finally, $R_C$ has been introduced in~\cite{Tripathi1999} to apply the same formalism also for the case of light nuclei. This parameter is different from 1 only for the reactions p$+$d, p$+^3$He and p$+^4$He, for which $R_C$ is assumed to be 13.5, 21 and 27, respectively (values taken from Table 2 of ~\cite{Tripathi1999}).

\item[CROSEC]

{In this approach we make use of the CROSEC fortran code\footnote{The original CROSEC code is available as a fortran file of the public \href{http://galprop.stanford.edu}{GALPROPv54} package.} authored by Barashenkov and Polanski.} 
CROSEC allows to compute the cross sections (total, nonelastic, elastic) for interactions from MeV/n to TeV/n energies. 
The cross sections have been evaluated by fitting parametric models to measurements at energies above several MeV/n (a detailed list of these measurements, together with the respective references, can be found at \href{http://www.oecd-nea.org/dbdata/bara.html}{http://www.oecd-nea.org/dbdata/bara.html}).

\end{description}

{As far as protons are concerned, the transport equation has to take into account the inelastic interactions with the ISM in which the particles lose a substantial part of their energy via pion production.  
Pion production is also responsible for gamma-ray emission, which allows to indirectly infer the proton Galactic distribution~\cite{1988A&A...207....1S,1996A&A...308L..21S,Yang:2016jda}. 

We consider here an additional approach to this process with respect to the one presented in our previous paper \cite{2017JCAP...02..015E}: We treat it by splitting the proton propagation in two steps: We first propagate the ``primary'' population with the pion production modeled as an inelastic loss term, and then  consider a ``secondary'' proton component made up by the particles that already suffered an inelastic collision and lost part of their energy. Both approaches are available in the code to test the differences in the propagated fluxes.}

\begin{figure}
\begin{center}
\includegraphics[width=0.49\columnwidth]{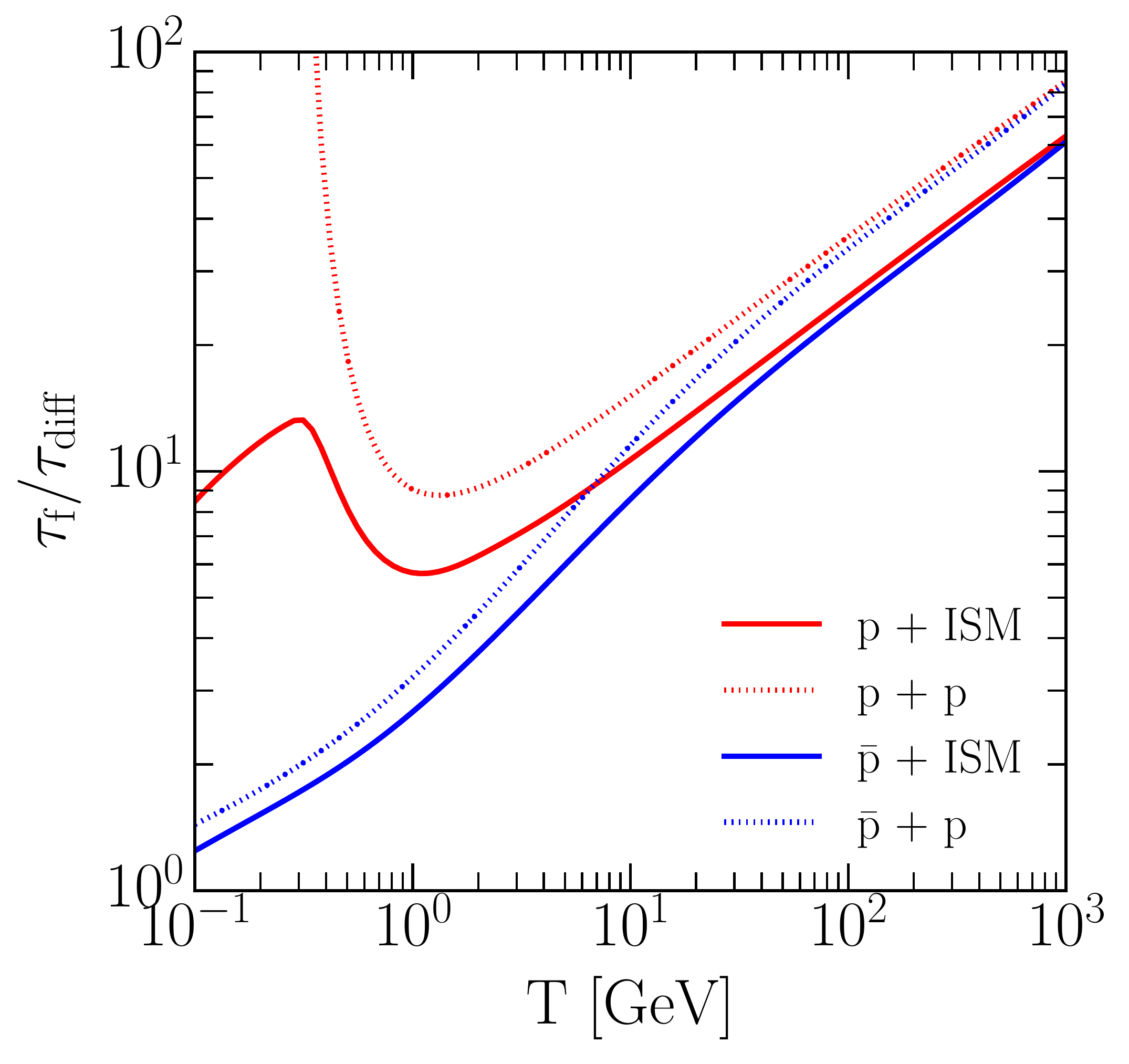} 
\hspace{\stretch{1}}
\includegraphics[width=0.49\columnwidth]{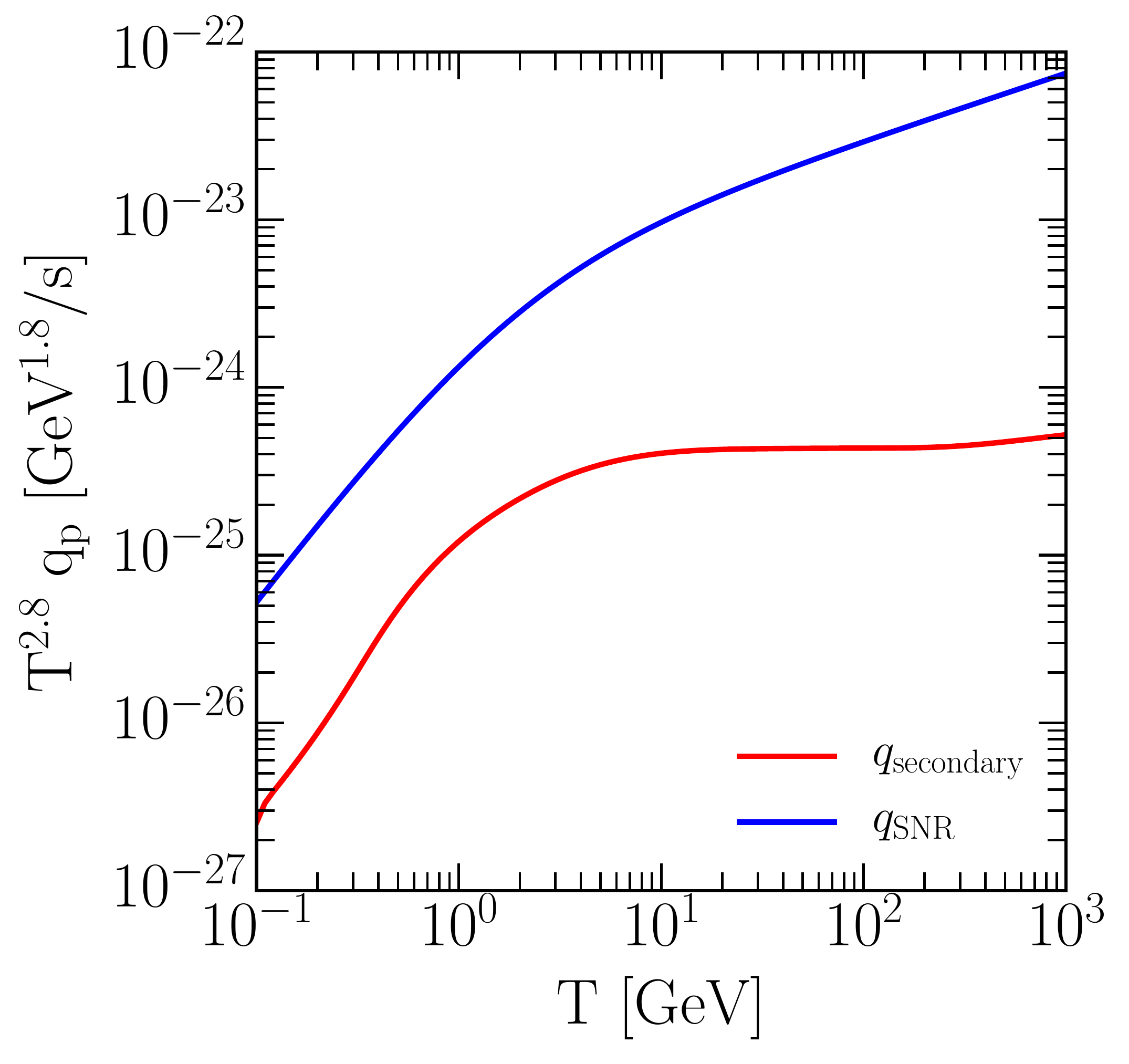} 
\end{center}
\caption{Left panel: Ratio between the inelastic and diffusion timescales for protons (blue lines) and antiprotons (red lines). {For antiprotons also annihilation is taken into account.}  {The propagation parameters are the same as in footnote 4.} The dotted lines correspond to the case in which only ISM hydrogen is considered. Right panel: The secondary proton production rate as a function of energy (red line) is compared with the primary source term (blue line). }
\label{fig:protons}
\end{figure}

As said, we assign a fragmentation rate to primary component as:
\begin{equation}
\frac{1}{\tau^{\rm f}_p(T)} = c n_{\rm H} \beta(T) \, \left[
{\sigma_{\rm pp}(T) \, + f_{\rm He} \, \sigma_{\rm pHe}(T)}  
\right] 
\end{equation}
where the cross section for the pp reaction has been evaluated in~\cite{2014PhRvD..90l3014K} from a fit to the total and elastic cross sections measurements:
\begin{equation}
\sigma_{\rm pp}(x = T/T_{\rm th}) = \left( 30.7 - 0.96 \log x + 0.18 \log^2 x \right) \times \left( 1 - x^{1.9} \right)^3 \,\, \text{mb}
\end{equation}
where $T_{\rm th} = 2 m_\pi + m_\pi^2 / 2 m_p \sim 0.2797$~GeV is the threshold kinetic energy.
The inelastic pHe cross section is taken from the parametrisation of Tripathi99 as discussed above. 
In the right panel of figure~\ref{fig:protons} we show that even when including the contribution of interstellar helium as target, the fragmentation timescale is always much longer than the escape time in the Galaxy.

The source term for the proton secondary component can be obtained from equation~(\ref{eq:secondary}) as:
\begin{equation}\label{eq:protonsec}
Q^{\rm sec}_p(T) = c n_{\mathrm{H}} \int_0^\infty dT' \beta(T') 
\left[ \frac{d\sigma_{\rm pp}}{dT}(T,T') + f_{\rm He} \frac{d\sigma_{\rm pHe}}{dT}(T,T') \right] 
N_{\rm p}(T') 
\end{equation}
where the differential cross sections can be further expanded as
\begin{equation}
\frac{d\sigma_i}{dT}(T,T') = \sigma_i^{\rm in}(T') \frac{dN}{dT}(T,T') \sim \frac{\sigma_i^{\rm in}(T')}{T'}
\end{equation}
using for the distribution of protons after scattering the approximation $dN/dE \sim 1/T'$ as in~\cite{1983JPhG....9..227T}\footnote{Notice that $dN/dE=dN/dT$}.

The relative contribution of the secondary source term to the total galactic population is shown, in comparison to the primary term, in the right panel of figure~\ref{fig:protons}.  
We compute the secondary source term by assuming that the proton spectrum $N_p$ in equation~(\ref{eq:protonsec}) is the proton local interstellar spectrum obtained by means of the analysis discussed in~\cite{2016A&A...591A..94G}. 
From figure~\ref{fig:protons}, one can deduce that at energies below $\sim 10$~GeV, secondary production contributes less than $\sim 10$\% to the total source term, and it becomes irrelevant at higher energies. 

\section{Nuclear decays}\label{sec:decays}

In case of unstable particles the radioactive decay rate is determined by
\begin{equation}
\frac{1}{\tau^{\rm r}_i(T)} = \frac{1}{\gamma(T) \tau^{\rm 0}_i}
\end{equation}
where $\tau^0$ is the isotope lifetime at rest\footnote{Lifetime is related to half-life in the following way: $\tau_{1/2}= \tau_0 \ln 2$.}. 
Among all the radioactive nuclei, those with lifetimes of the order of the propagation time (e.g., $^{10}$Be) may bring interesting information about the primary residence time in the Galaxy~\cite{Berezinskii1990}.

If the decay lifetime at rest is much shorter than the minimum propagation timescale (but larger than $\sim 1$~ms) the isotope may be assumed to decay immediately after production.
In this case we simply add the contribution of all the ``intermediate lived'' parent nuclei to the corresponding secondary source term (see discussion in section~\ref{sec:frag}).

Similarly to the other codes as GALPROP or USINE, in \dragon~the CR abundances are evaluated from the heaviest to the lightest element. This means that, when modeling spallation reactions, the code creates a network where the different particle species are sorted in descending order according to their atomic number. It is clear that such an approach cannot be adopted when dealing with radioactive species that decay via $\beta^-$ emission, since a decay of this kind causes an increase in the atomic number.

We overcome this problem by inverting the order of the two nuclei, such that we evaluate the lighter first and then we use it as an input for the decayed one. We repeat the evolution of the two coupled species twice, hence we are able to take into account the spallation of the heavier species into the lighter one. 
 
We do not implement EC capture and decay in the current version of the code. Indeed, we tested with the public version of \galprop~that considering these nuclei as stable does not impact on our results on Li, Be and B local fluxes for more than 1\%. 
We plan to implement this process in a future work focused on the sub-Fe/Fe ratio.

Decay rates and modes are taken from the {National Nuclear Data Centre Database}\footnote{\url{http://www.nndc.bnl.gov}} and are {reported the table available at \url{https://github.com/cosmicrays}} for the isotopes included in~\dragon.

\section{Fragmentation processes}
\label{sec:frag}

\begin{figure}[t]
\begin{center}
\includegraphics[width=0.48\columnwidth]{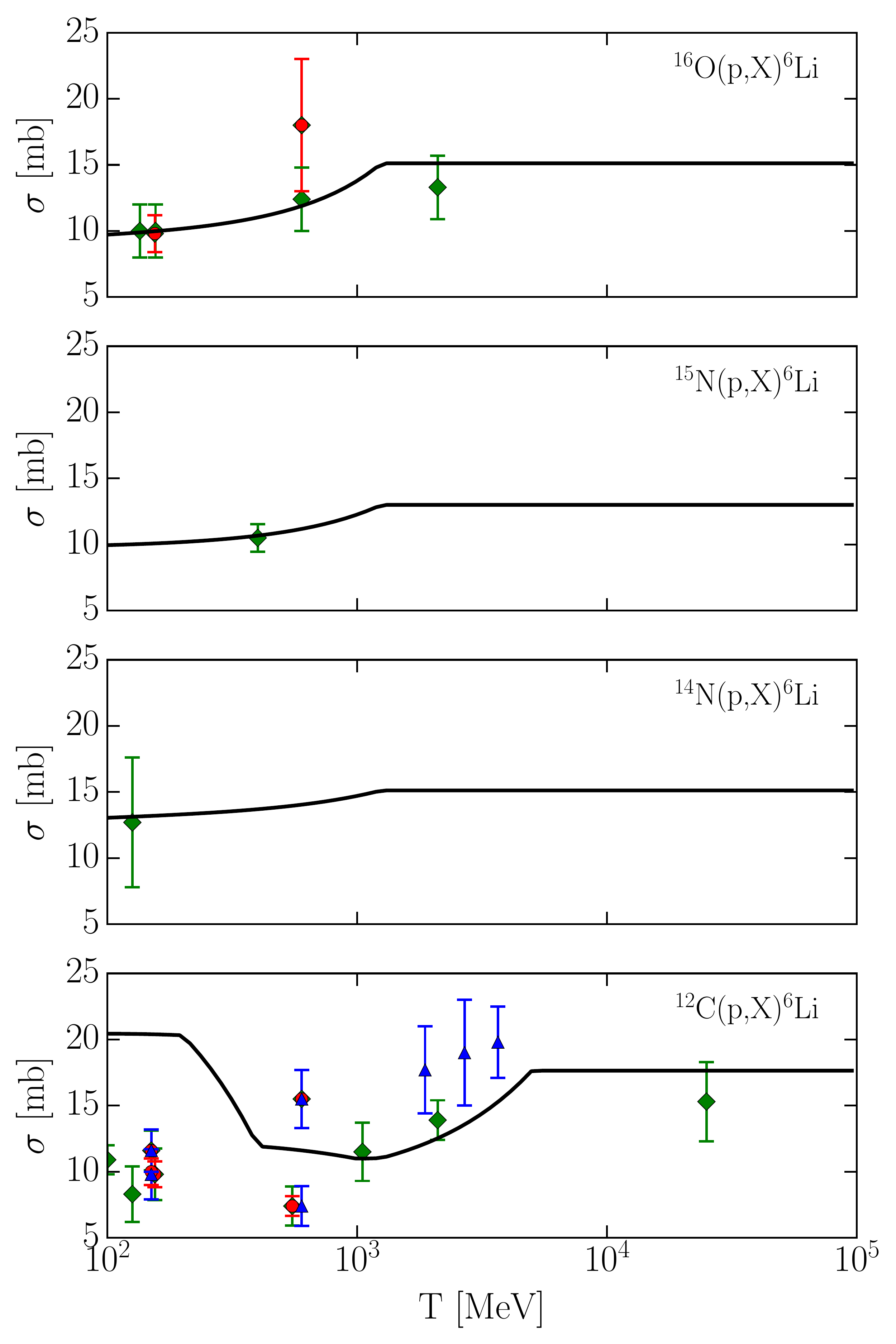} 
\includegraphics[width=0.48\columnwidth]{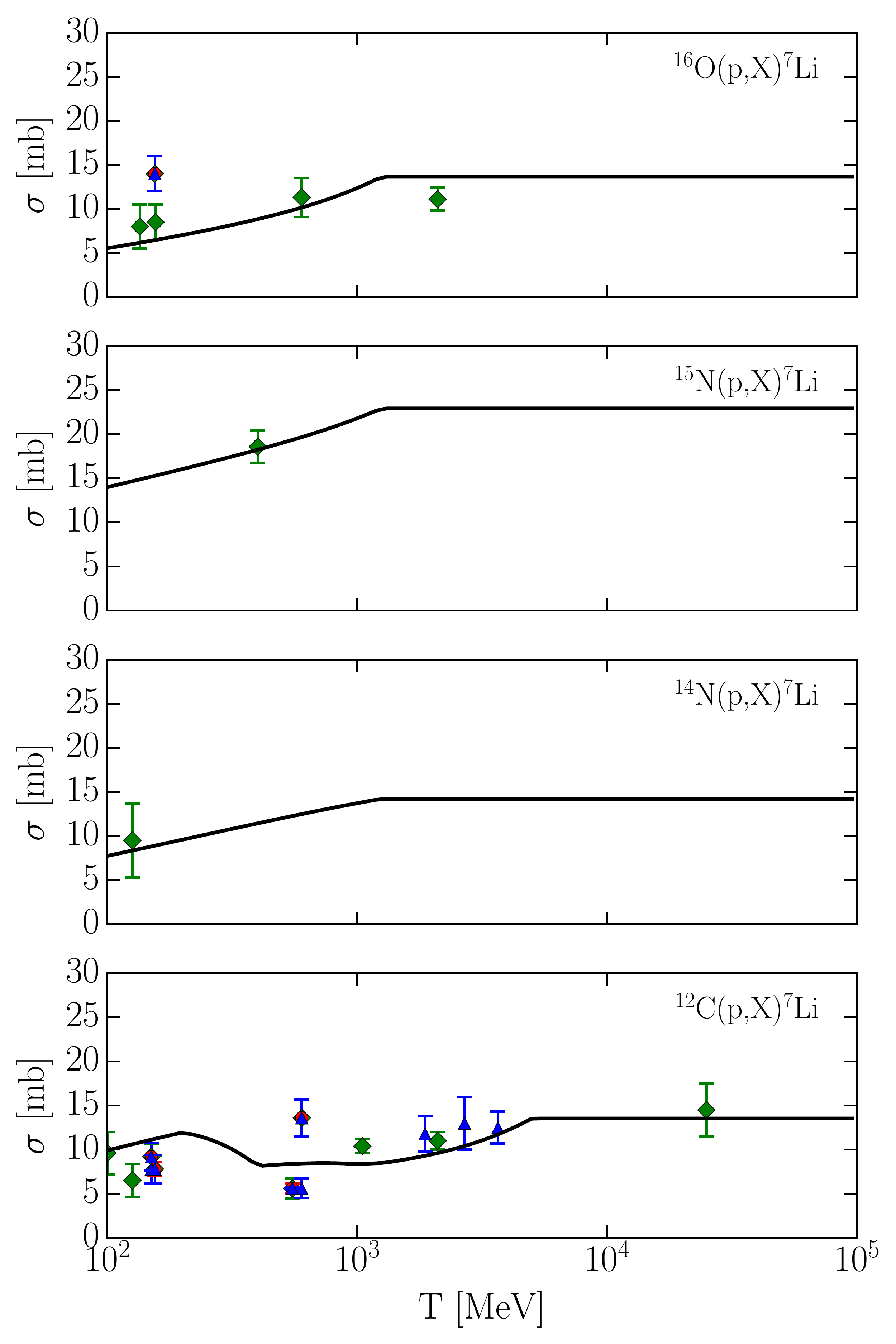} 
\end{center}
\caption{Comparison between fragmentation model in \dragon~and measurements for the main channels fragmenting in $^6$Li and $^7$Li. Measurements shown with green (diamond) points are taken from \galprop~database, with red (circle) points from \exfor~and blue (triangle) from~\cite{2000JPhG...26.1171K,2002JPhG...28.1199K}.}
\label{fig:Li}
\end{figure}

\begin{figure}[t]
\begin{center}
\includegraphics[width=0.48\columnwidth]{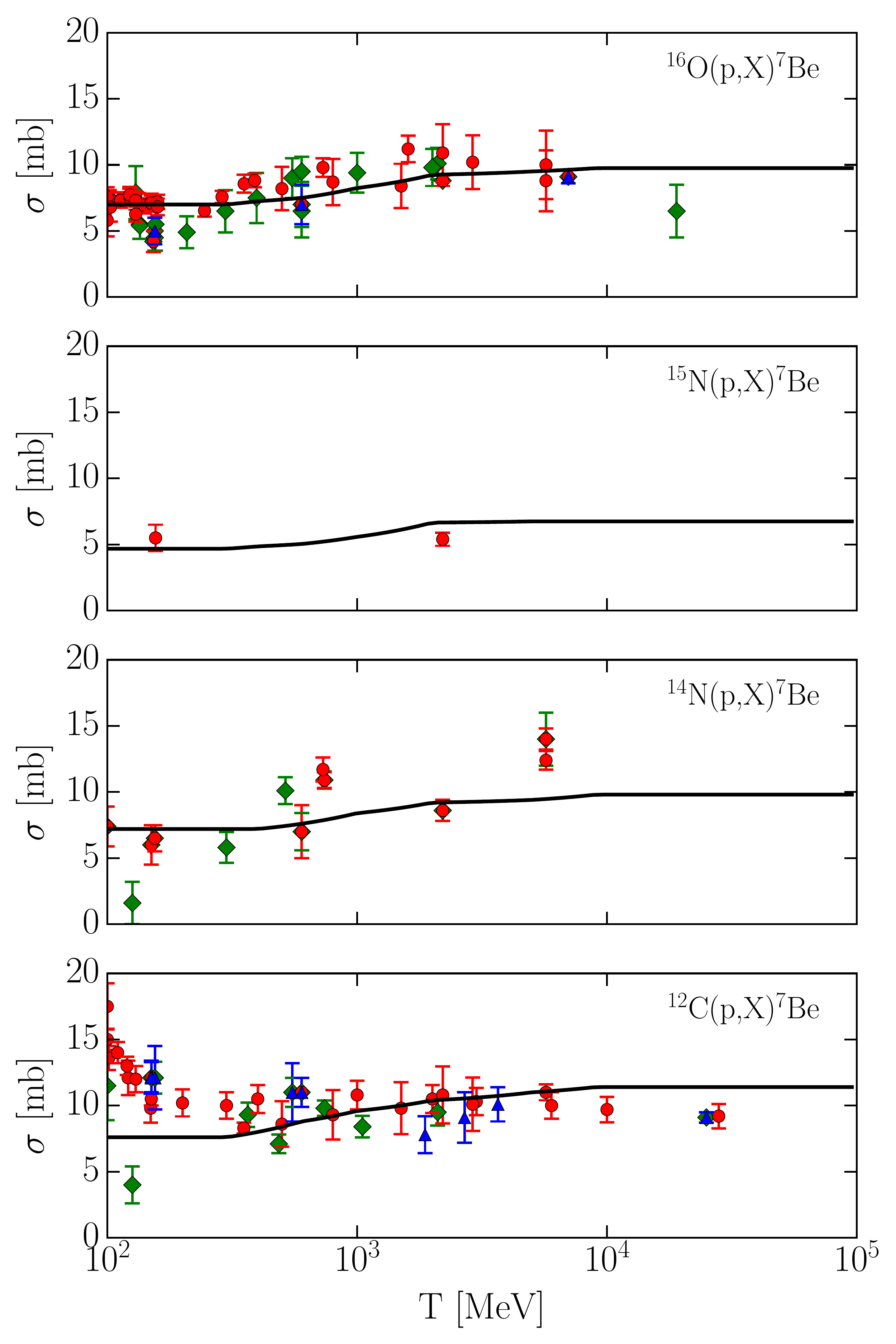} 
\includegraphics[width=0.48\columnwidth]{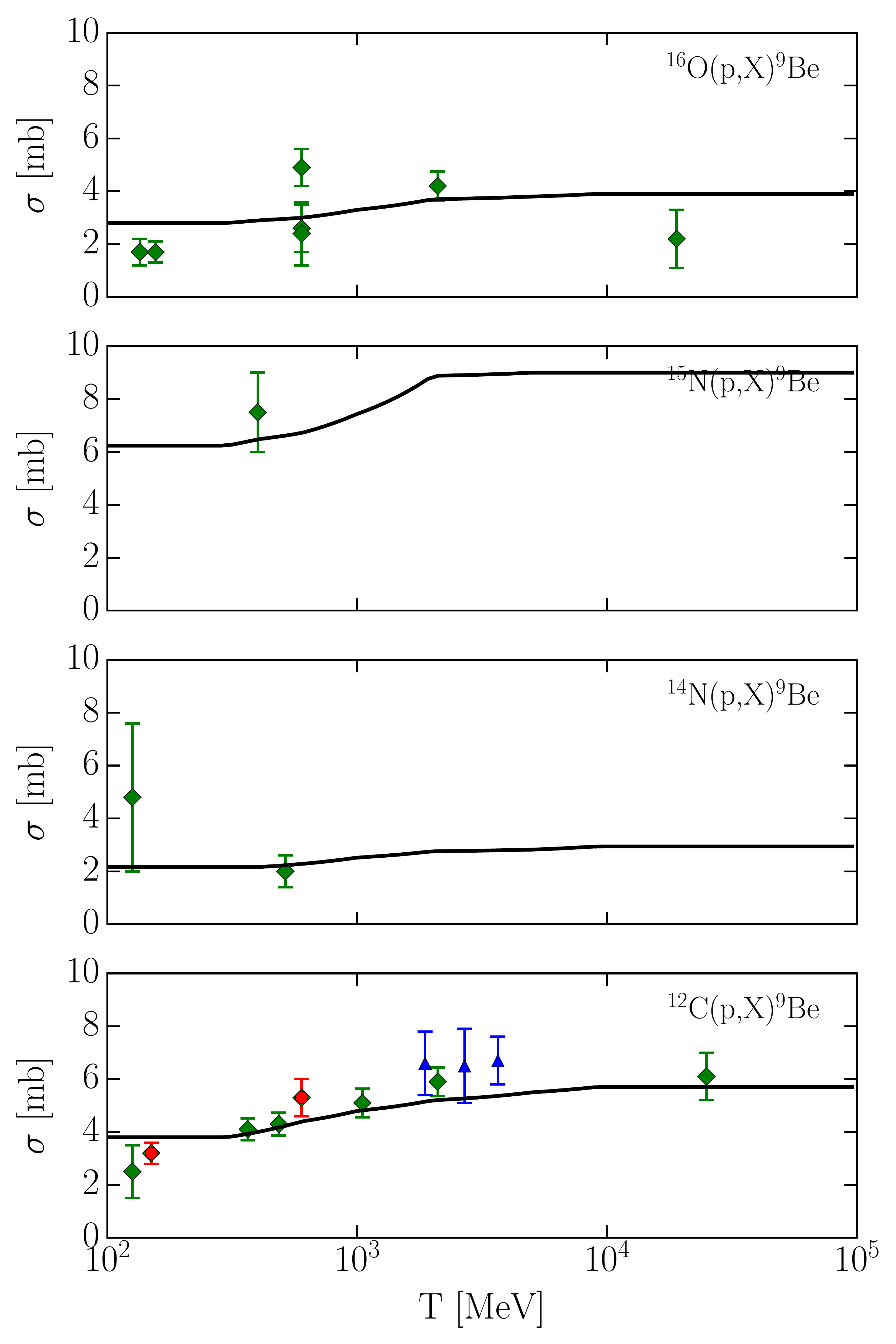} 
\end{center}
\caption{The same as in figure~\ref{fig:Li} for $^7$Be and $^9$Be.}
\label{fig:Be1}
\end{figure}

\begin{figure}[t]
\begin{center}
\includegraphics[width=0.48\columnwidth]{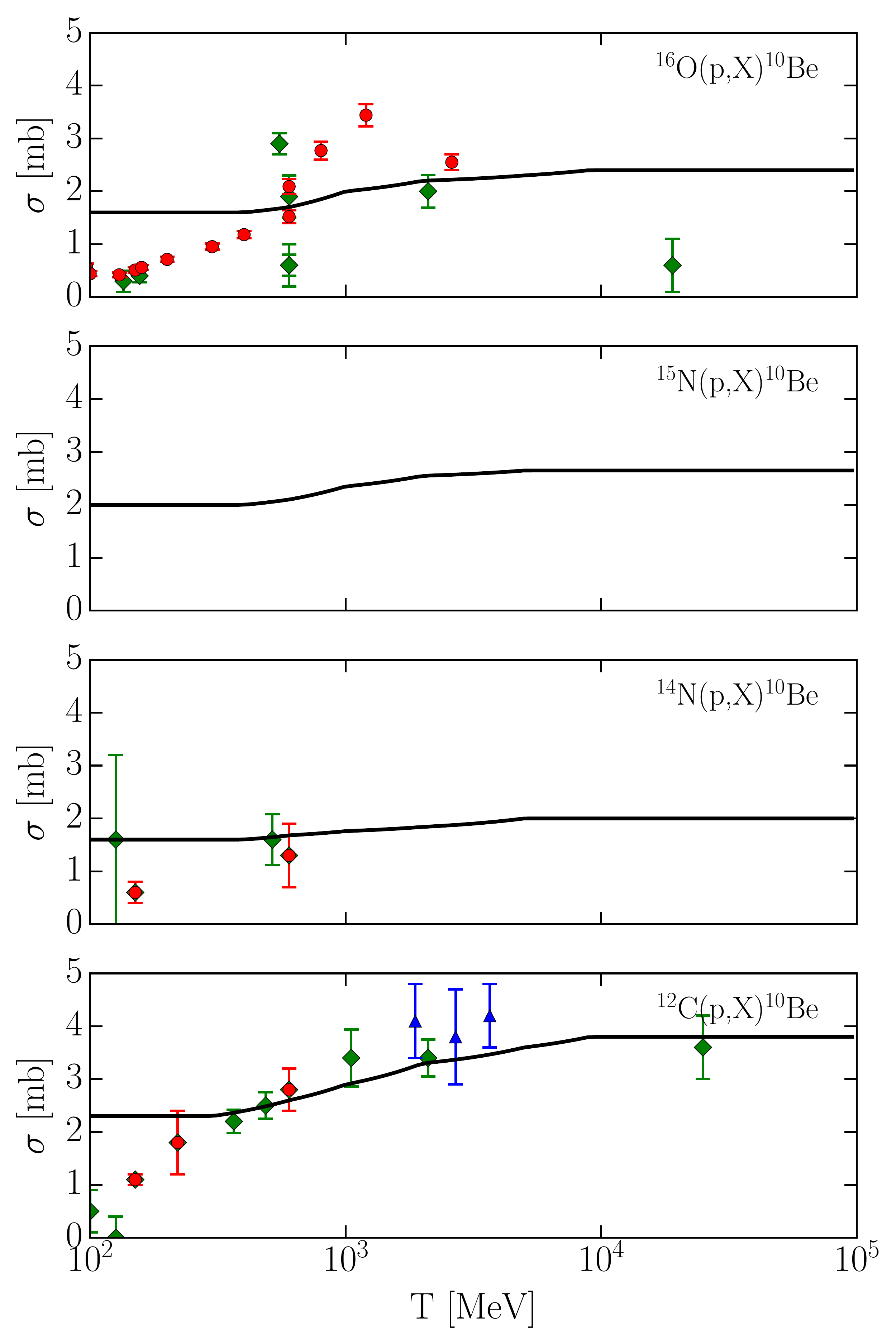} 
\includegraphics[width=0.48\columnwidth]{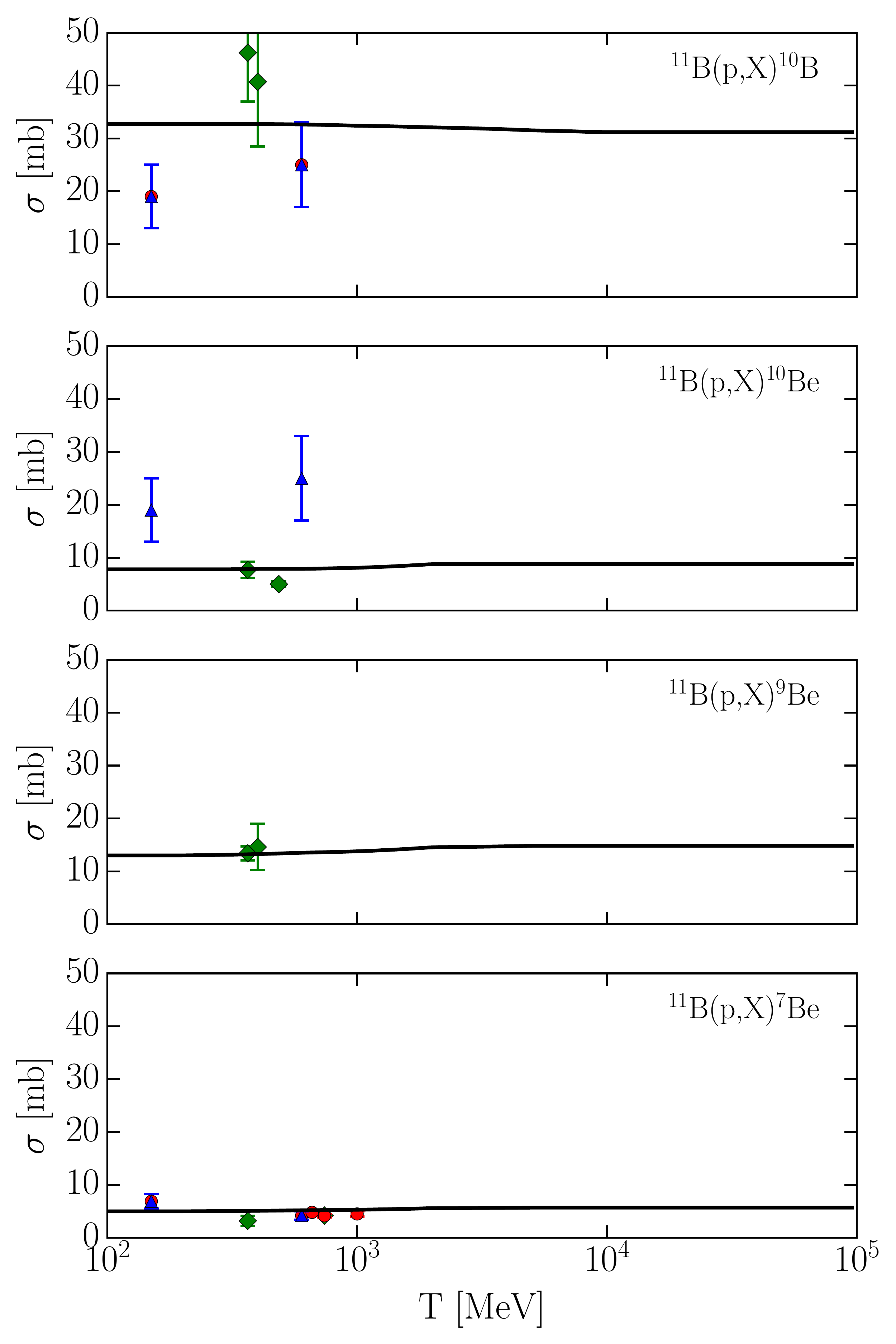} 
\end{center}
\caption{The same as in figure~\ref{fig:Li} for $^{10}$Be (left) and for different channels assuming $^{11}$B as projectile.}
\label{fig:Be2}
\end{figure}

\begin{figure}[t]
\begin{center}
\includegraphics[width=0.48\columnwidth]{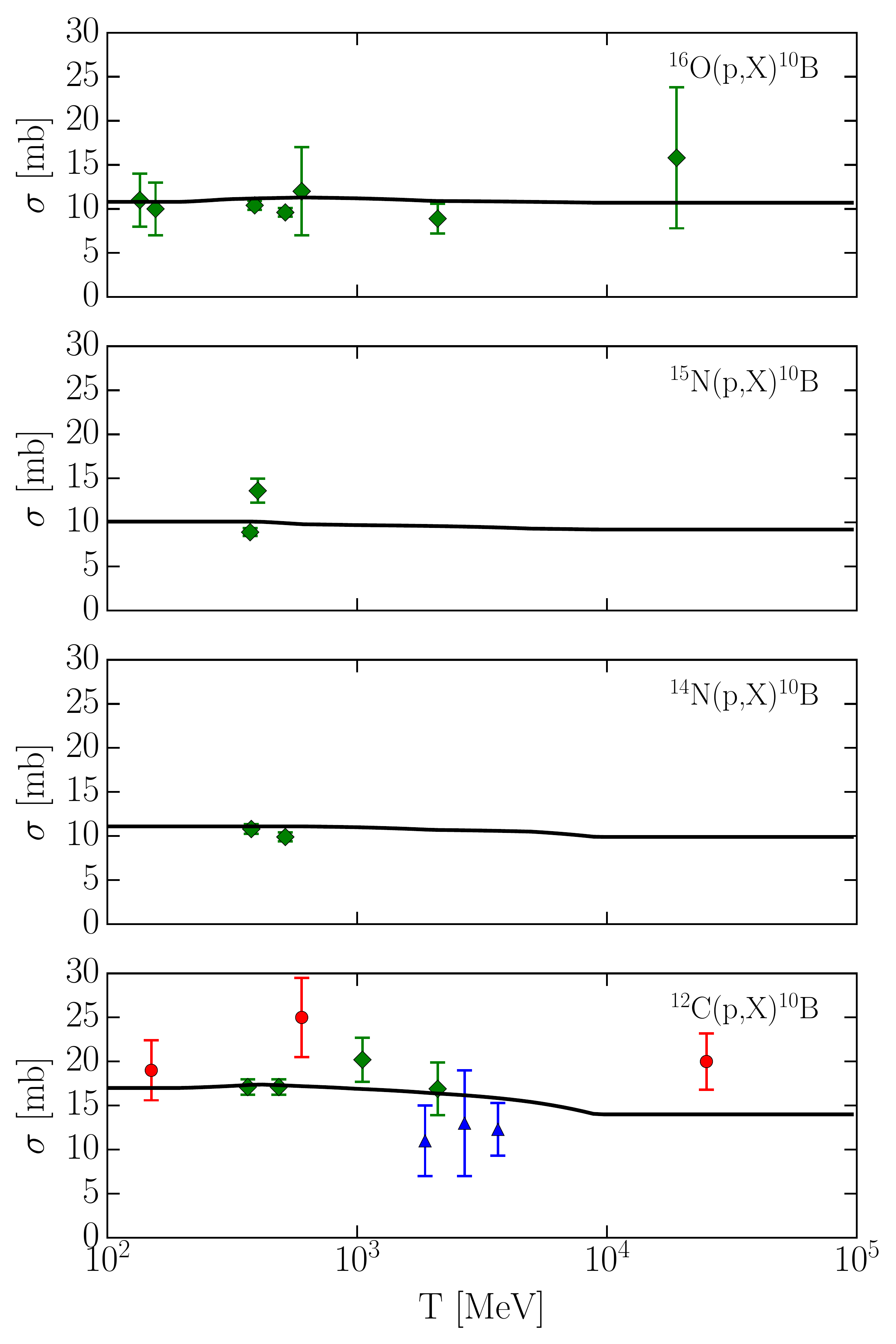} 
\includegraphics[width=0.48\columnwidth]{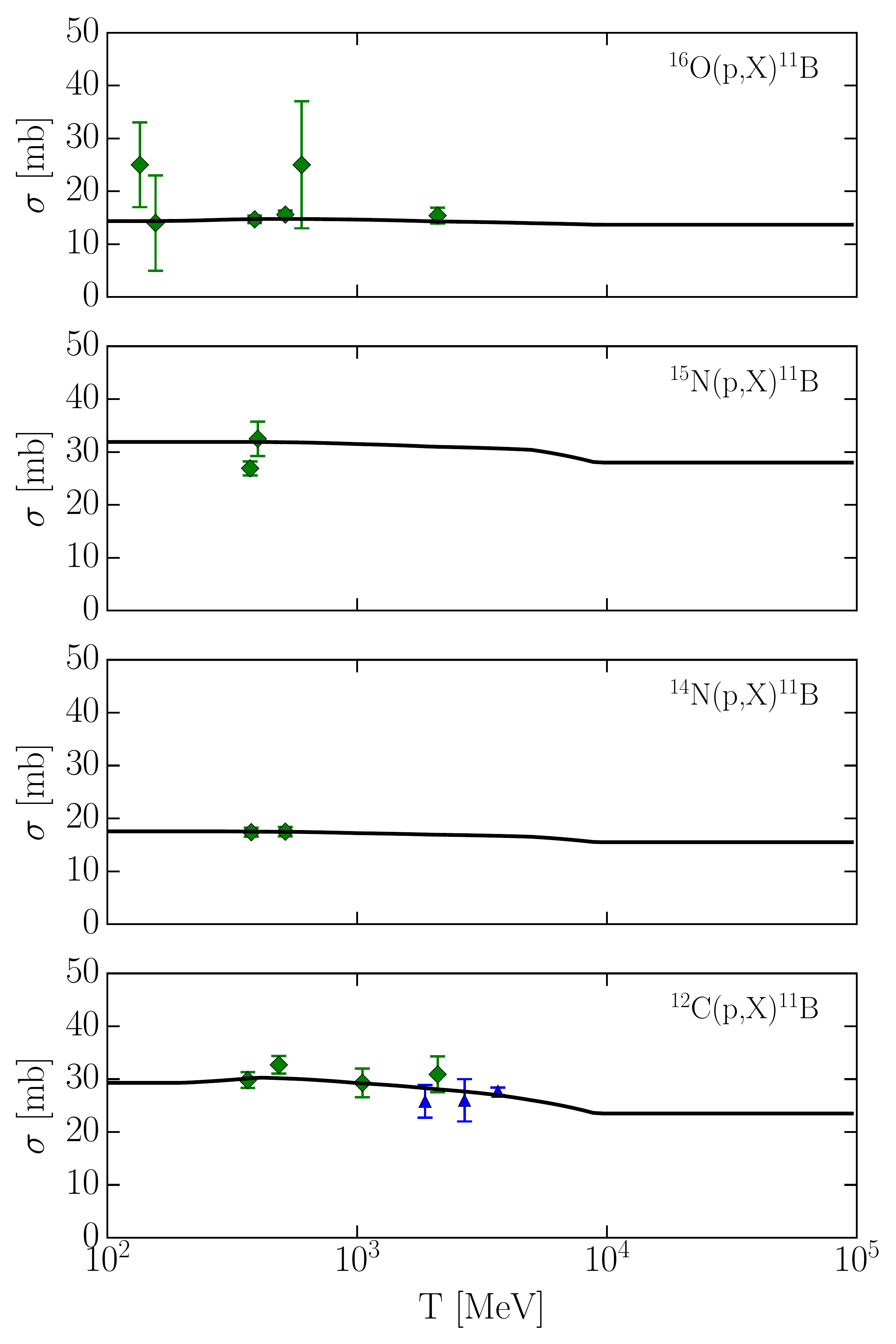} 
\end{center}
\caption{The same as in figure~\ref{fig:Li} for $^{10}$B and $^{11}$B.}
\label{fig:Boron}
\end{figure}

Spallations (or fragmentations) are processes in which the interaction of a CR nucleus with a light target (in our case, interstellar proton or helium nuclei) causes the emission of a large number of hadrons (mostly neutrons) or heavier fragments~\cite{1947PhRv...72.1114S}.

We assume conservation of kinetic energy per nucleon in all fragmentation processes, as typically done in both numerical and semi-analytic approaches \footnote{see~\cite{maurin:tel-00008773} for an attempt to quantify the accuracy of this approximation in the case of B/C}:
\begin{equation}
\frac{d\sigma_{{\rm k},j \rightarrow i}}{dT}(T,T') = \tilde{\sigma}_{{\rm k},j \rightarrow i} \delta(T-T')
\end{equation}
where the terms $\tilde{\sigma}_{{\rm k},j \rightarrow i}$ represent the spallation cross sections associated to the production of species $i$ from the projectile $j$ impinging on the gas component $\mathrm{k}$.

Under this simplifying assumption, the source term associated to the secondary products of spallation can be written as:
\begin{equation}
\Gamma^{\rm s}_{j \rightarrow i}(T) = \beta(T) c n_{\rm H} N_j(T) 
\left[ 
\tilde{\sigma}_{{\rm H},j \rightarrow i}(T) + f_{\rm He} \tilde{\sigma}_{{\rm He},j \rightarrow i}(T) 
\right]
\end{equation}

Adequate assessment of secondary production (and consequently of the Galactic grammage) requires accurate methods aimed at computing these quantities. 

At present, there is no accurate theory that predicts the spallation cross sections for all collision pairs and energies of interest in CR physics.

However, some scaling relations have been found, for instance: the isoscaling phenomenon, the m-scaling\footnote{where $m$ is defined, in terms of the atomic number, mass number and neutron number of the fragment, as $m = (N-Z)/ A$}, and the isobaric ratio difference scaling (see, e.g.,~\cite{2016PrPNP..91..203B}). 
These scaling phenomena for fragments reflect general properties of the reaction, and play a crucial role in the determination of semi-empirical formul\ae.

At energies of interest for CR studies, the two relevant results on this line of research are ({see~\cite{maurin:tel-00008773}  and~\cite{2018arXiv180304686G} for a longer discussion}):
\begin{itemize}
\item In~\cite{1973ApJS...25..315S,1973ApJS...25..335S}, Silberberg and Tsao took advantage of the observed regularity in the mass difference between target and daughter nuclei and on the ratio between the number of neutrons and protons in the daughter nucleus to provide a fast estimator of spallation cross sections up to $Z = 83$.

\item In the 1990s, Webber and coworkers obtained many new data from various targets, allowing them to develop a different approach fully based upon experimental regularities~\cite{1998PhRvC..58.3539W,1998ApJ...508..940W,1998ApJ...508..949W,2003ApJS..144..153W}. 

\end{itemize}

In \dragon~for each species that receives a secondary contribution, the spallation cross sections are computed by following this algorithm:
\begin{itemize}
\item We first search for the spallation channel in the cross section table provided by Webber, and corresponding to the results reported in~\cite{2003ApJS..144..153W}. These tables provide {\it cumulative} cross-sections, i.e., they already include the contribution that comes from intermediate lived parent nuclei.

\item If the channel is not present in the Webber table, we use the {\tt WNEWTR} (version 1983) code\footnote{The original code by Webber is available as a FORTRAN file within the \href{http://galprop.stanford.edu}{GALPROPv54} package.} to compute the direct channel. Whereas data are available, we tune the overall normalization of the cross section on the existing measurements in the energy range of interest $0.1 \le T \le 100$~GeV/n. This is different than the {\tt GALPROP} approach where a combination of methods is adopted, partly based on their fits of a compilation of cross-section measurements and code evaluations, partly based on interpolation of measurements based on the Webber and Silberberg models~\cite{2001ICRC....5.1836M,2003ICRC....4.1969M,2003ApJ...586.1050M}. 

Figures~\ref{fig:Li}, \ref{fig:Be1}, \ref{fig:Be2}, \ref{fig:Boron} show the agreement between the model and measurements for the most relevant channels.
The measurements shown in these plots are mainly taken from:
\begin{description}

\item[EXFOR] (Experimental Nuclear Reaction Data)\footnote{\url{https://www-nds.iaea.org/exfor/exfor.htm}} which is an extensive database containing experimental data, their bibliographic information, experimental information and source of uncertainties~\cite{2014NDS...120..272O}.
Additional information about the experimental setup (most of them provided by the authors) is available through the web interface allowing to perform a reasoned comparison with theoretical predictions.

\item[GALPROP] includes a database of cross-sections measurements\footnote{In the file 
\url{isotope_cs.dat}} for most of the relevant channels in the energy range from $\sim$~MeV/n to $300$~GeV/n. 

\end{description}

Additionally, we make use of the {\tt WNEWTR} code to compute the contribution of all the intermediate lived parent nuclei (lifetime shorter than $1$~hour and longer than $1$~ms) as in~\cite{2001ApJ...555..585M}:
\begin{equation}
\tilde{\sigma}^{\rm effective}_{{\rm H},j \rightarrow i} = \tilde{\sigma}_{{\rm H},j \rightarrow i} + \sum_{X} \tilde{\sigma}_{{\rm H},j \rightarrow X} \mathcal{B}(X \rightarrow i)
\end{equation}
where $\mathcal{B}$ is the branching ratio of the channel $X \rightarrow i$. 
In fact, we extend the decay chain up to 3 generations of parent nuclei. As a comparison, \galprop~accounts up to 5 generations of the decay products~\cite{2001ICRC....5.1942S,2002ApJ...565..280M}.

\item For secondary Li production, we use the parametrization provided by Silberberg and Tsao, since Webber at al.~parameterisations do not include Li production cross sections.
In this case, we compute the cross sections by using the {\tt YIELDX} (version 1999) code\footnote{The original code by Silberberg and Tsao is available as a fortran file of the \href{http://galprop.stanford.edu}{GALPROPv54} package.}.

\item For nuclear fragmentation cross sections in $^2$H, $^3$H, and $^3$He we adopt the fitting functions reported in \cite{2012A&A...539A..88C}.
In particular an useful factorization of the proton-induced reactions is given in their section B.2.3, and a set of formul\ae~to compute the factors are provided. 

\end{itemize}

\begin{figure}
\begin{center}
\includegraphics[width=0.49\columnwidth]{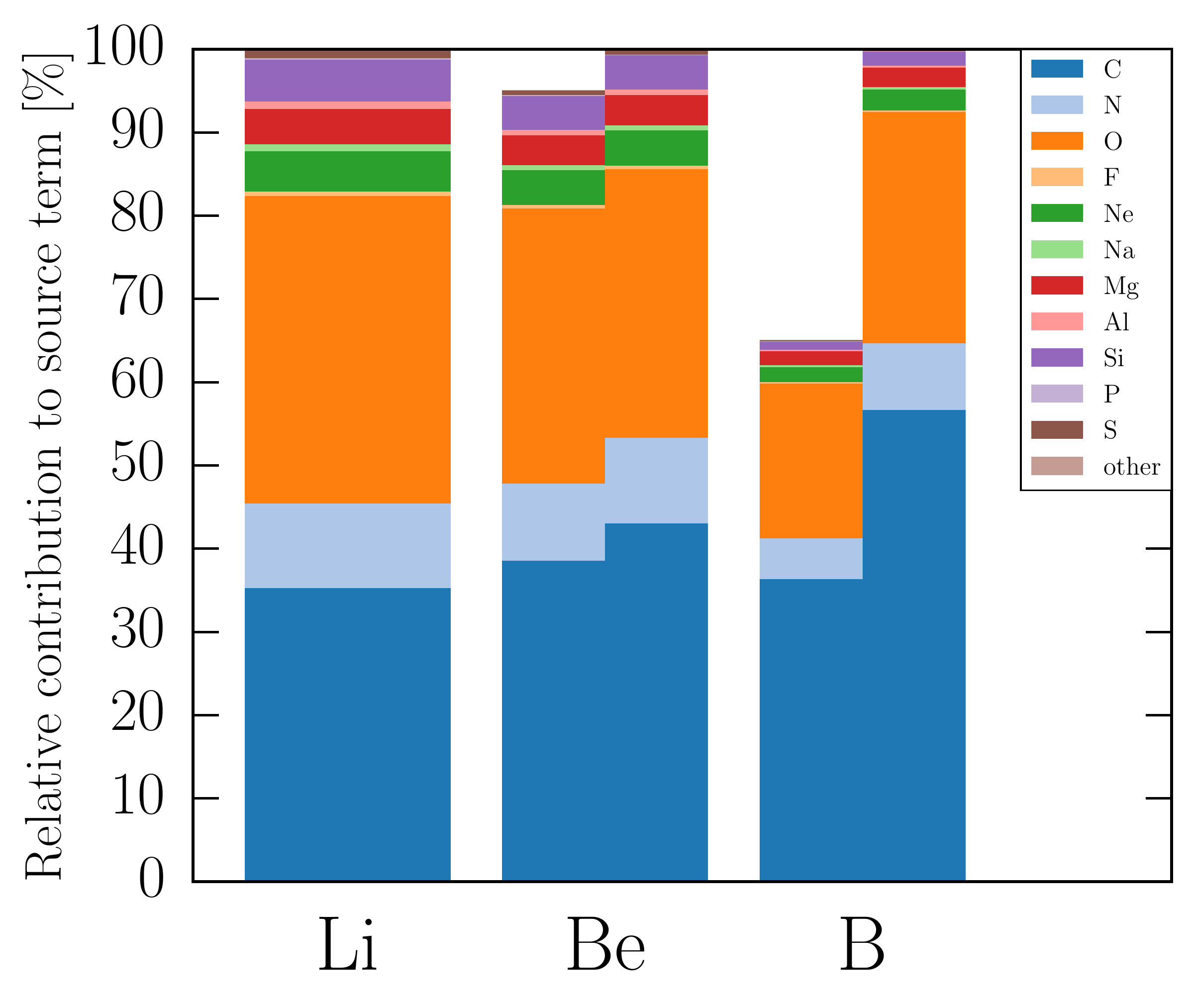} 
\includegraphics[width=0.45\columnwidth]{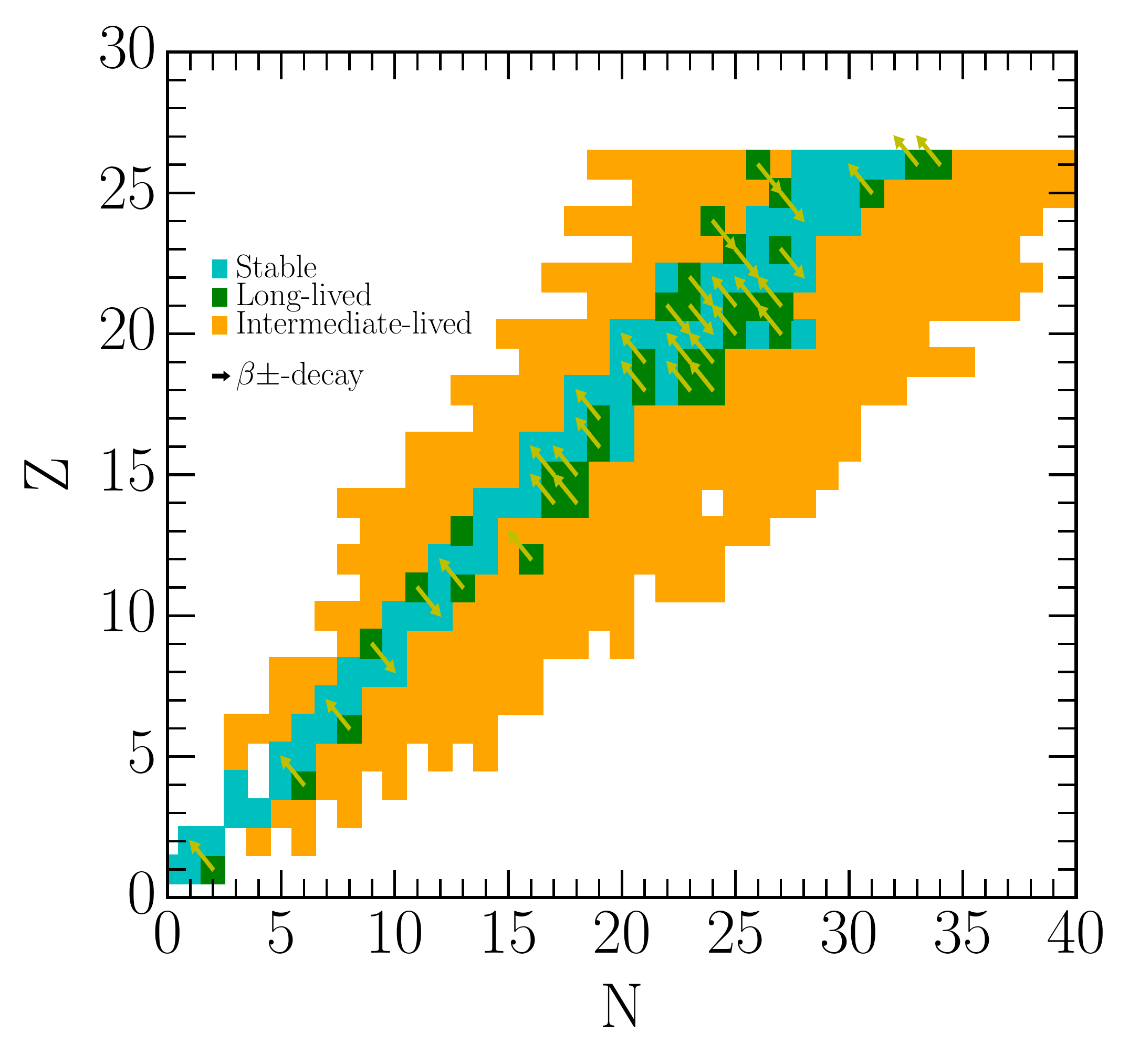} 
\end{center}
\caption{Fractional contribution to the secondary source term for Li, Be, B from heavier species at $10$~GeV/n. The bars on the left are obtained without including the contribution to the cross-sections from intermediate nuclei.}
\label{fig:libeb}
\end{figure}

To account for interstellar helium, an empirical formula for the ratio of partial cross sections as a function of the energy and the charge change is provided by~\cite{1988PhRvC..37.1490F}.

\begin{figure}
\begin{center}
\includegraphics[width=0.48\columnwidth]{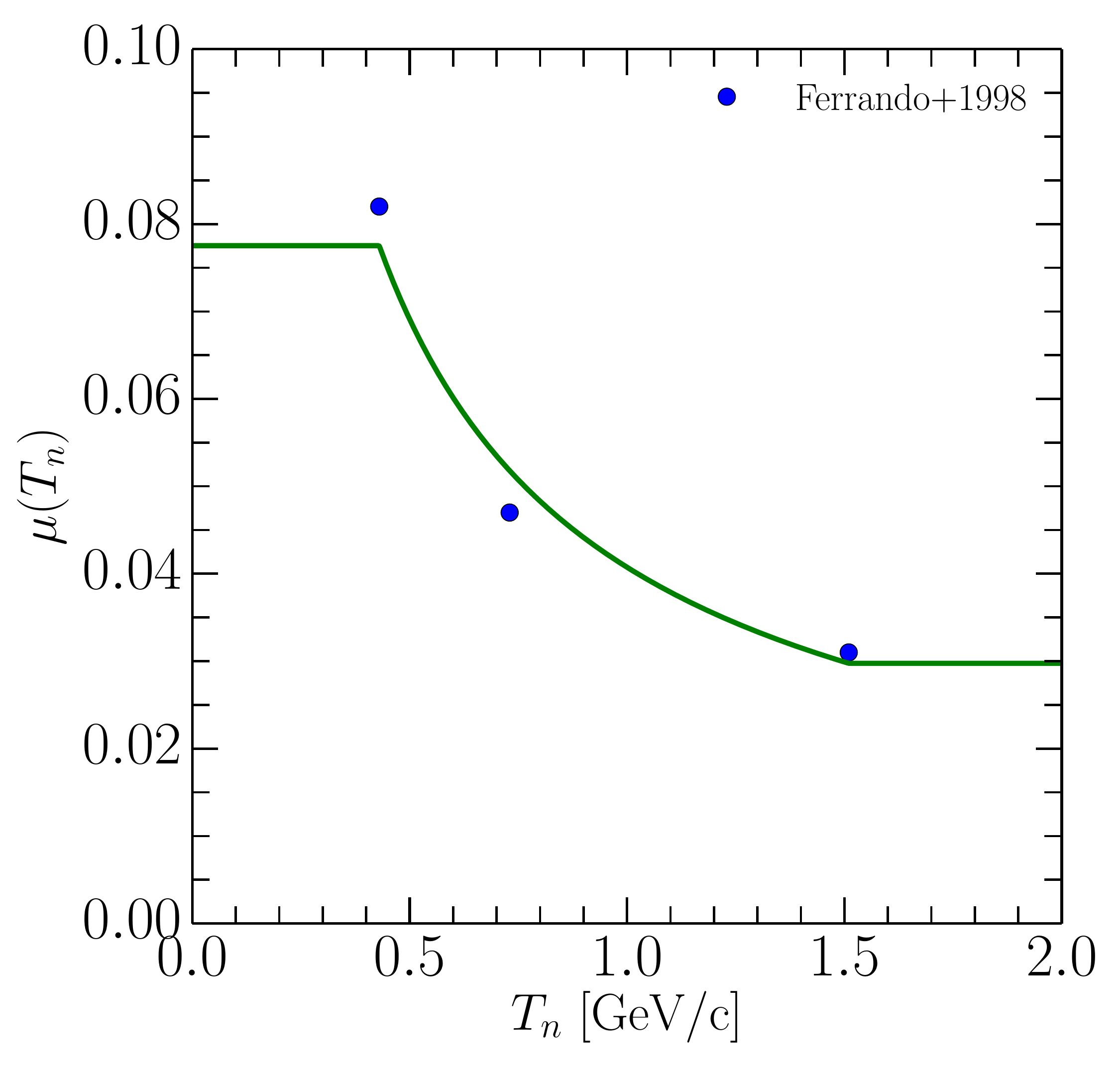} 
\hspace{\stretch{1}}
\includegraphics[width=0.48\columnwidth]{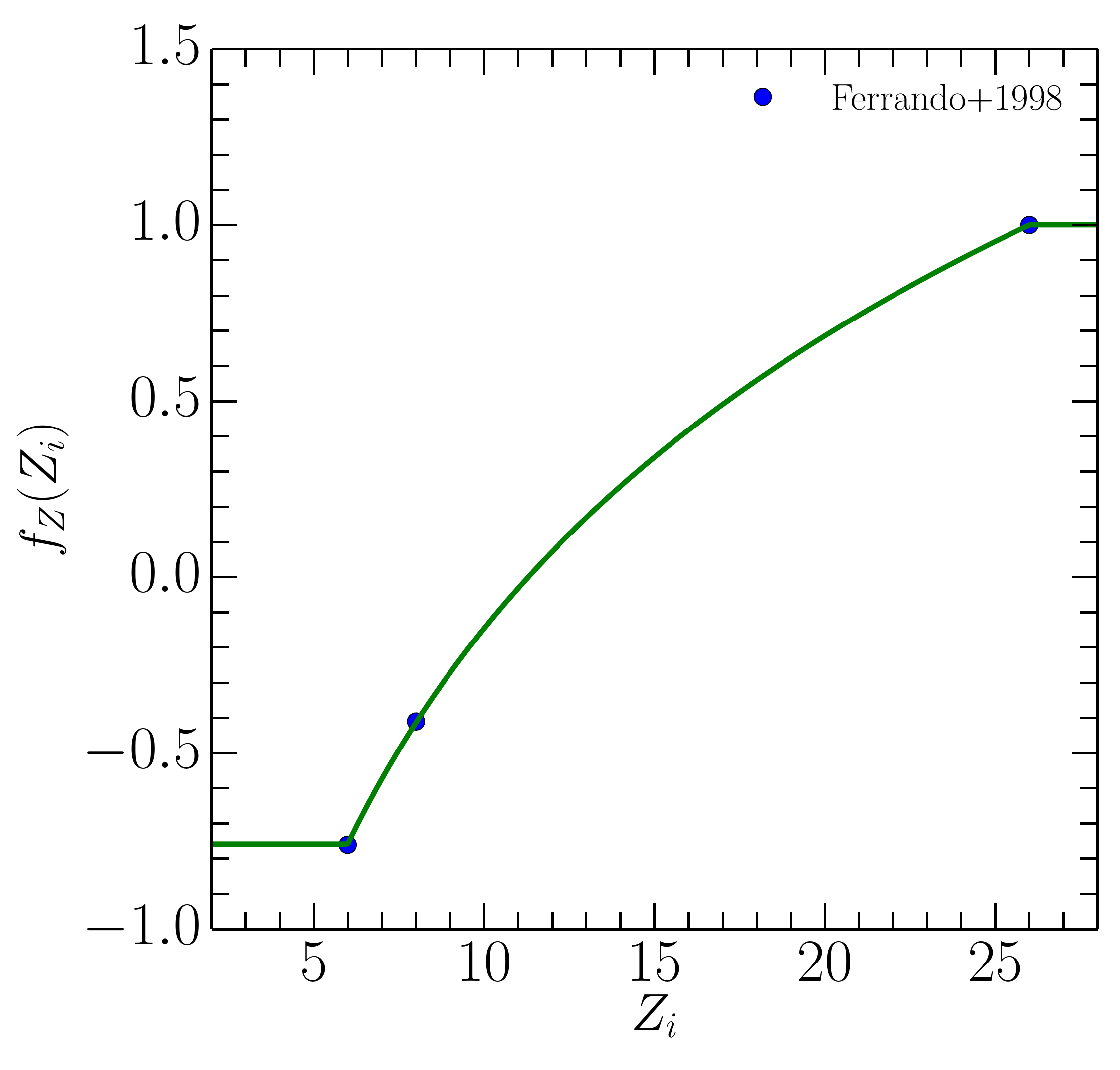} 
\end{center}
\caption{Second-order polynomial fit of the functions in equation~\ref{eq:He2H}.}
\label{fig:He2H}
\end{figure}

In this approximation:
\begin{equation}\label{eq:He2H}
\frac{\tilde{\sigma}_{{\rm He},j \rightarrow i}}{\tilde{\sigma}_{{\rm H},j \rightarrow i}} (Z_i, Z_f, T) = 
\exp \left[ \mu(T) | Z_i - Z_f - f_Z(Z_i) \delta(T)|^{1.43} \right]
\end{equation}
where each of the functions $\mu(T)$, $\delta(T)$ and $f_Z(Z_i)$ is measured for three values of energy or atomic number. 
To implement this prescription in \dragon, we fit the data with a second order polynomial and we assume that the functions are constants outside the data range as in figure~\ref{fig:He2H}.

In figure~\ref{fig:libeb}, we show the fractional contribution to the secondary source term for Li, Be, B at $10$~GeV/n from heavier CR parents.
The primary spectra that we use to compute the source terms have been obtained by fitting with a single power-law all the available measurements at $T \gtrsim 10$~GeV/n as obtained from CRDB\footnote{The database of charged cosmic rays available at \url{https://lpsc.in2p3.fr/cosmic-rays-db/}.}.
We conclude that to account for the 90\% of the total source term 
it is sufficient to start the nuclear chain from Mg.
The relative contribution of intermediate lived nuclei is also shown.
They account for $\sim 35$\% to the B source term, becoming less relevant for lighter elements.

The effect of cross section systematics was studied by~\cite{2010A&A...516A..67M}, who parameterised it in terms of a systematic shift with respect to the energy. 
They conclude that different choice of the production cross section model gives a 20\% of systematic uncertainty on the determination of transport parameters. 
With a similar approach but assuming anti-correlated modifications in the destruction and production cross sections, the authors of~\cite{2015A&A...580A...9G}  estimated a systematic uncertainty of $\sim 10$\% on the diffusion coefficient slope $\delta$ and of 40\% on the coefficient normalization, $D_0$.

In~\cite{2017arXiv170706917T}, a data-driven re-evaluation of isotopic cross section involving the production of B-Be nuclei is presented.
By comparing these with other sources of uncertainties, it has been found that these represent a major limiting factor for the interpretation of the recent B/C data from AMS. 
Hence, they stress once more the urgent need for establishing a program of cross section measurements at the $\mathcal{O}(100 \, \rm{GeV})$ (an experimental program to measure carbon-proton and oxygen-proton interactions at $13 A$ GeV/c at the CERN SPS has been proposed in~\cite{NA61}).

More recently, \cite{2018arXiv180304686G} has discussed in full detail how to rank the most important reactions for the production of light secondary CR nuclei and provided a comprehensive discussion about the impact on the CR flux determination of current and future experimental efforts aimed at measuring the relevant cross sections.

%% file: leptons.tex
\section{Secondary leptons}
\label{sec:leptons}

The differential cross section describing the production of secondary electrons and positrons, originated by CR species $j$ colliding on the ISM particle $k$, can be written as: 
\begin{equation}
\left.\frac{d\sigma}{dT}\right|_{\mathrm{k},j \rightarrow e^{\pm}} (T,T') = \sigma_{\mathrm{k},j}(T') \frac{dn}{dT}(T,T')
\end{equation}
where $T$ is the lepton energy, $T'$ represents the energy of the particle $j$, $\sigma_{\mathrm{k},j}(T')$ is the total inelastic cross section for the $j+\mathrm{k}$ reaction and $dn/dT$ is the energy distribution of the electrons and positrons produced in a single collision.

The production of electrons and positrons in spallation processes happens through the decay of charged or neutral pions ($\pi^0 \rightarrow e^+ + e^-$ or $\pi^\pm \rightarrow \mu^\pm + \nu_\mu \rightarrow e^\pm + \nu_e + 2\nu_\mu$), {which, in turn, are produced either directly, or through the decay of other particles, in a multi-step process}. 

If we denote $f = dn/dT$, we can write: 
\begin{equation}
f_{e,\mathrm{k}}(T, T') = \int_{T_{\pi,\mathrm{min}}}^{T_{\pi,\mathrm{max}}} dT_{\pi} f_{\pi,\mathrm{k}}(T_\pi, T') \, f_{e,\pi}(T, T_\pi) 
\end{equation}   
where $f_{e,\mathrm{k}}$ is the energy distribution of $e^{\pm}$ produced in $p$k collisions (where k = H,He), while: 
\begin{itemize}
\item{$f_{\pi,\mathrm{k}}$ is the energy distribution of the pions that are produced in $p$k collisions. This process involves non-perturbative effects and therefore cannot be computed exactly: one should use either semi-empirical or Monte Carlo methods.}
\item{$f_{e,\pi}$ is the energy distribution of $e^{\pm}$ produced by the pion decay. This is an electroweak process and can be computed exactly with analytical methods.}
\end{itemize}
From these considerations, we can clearly conclude that all the differences among leptonic cross sections models deal with the $f_{\pi,\mathrm{k}}$ term. 

While semi-empirical models have been widely used in the past (e.g., \cite{1986A&A...157..223D,1986ApJ...307...47D,Blattnig:2000zf}), these models have been showing several shortcomings in particular concerning diffractive processes or the violation of the Feynman scaling (for a detailed discussion see ~\cite{Kamae:2004xx}).

One can try to overcome these limitations by simulating lepton production via Monte Carlo event generators.
In \dragon~we implement two different models based on this approach:  

\begin{description}

\item[Kamae2006] This model~\cite{Kamae:2004xx,Kamae:2006bf} is a combination of Monte Carlo and semi-empirical methods. In particular, it employs the parameterization taken from Blattnig et al.~\cite{Blattnig:2000zf} to model non-diffractive processes at low energy ($T_p \le 52.6$ GeV), the Monte Carlo event generator {\tt Pythia} 6.2 to account for for non-diffractive high-energy ($52.6~\mathrm{GeV} \le T_p \le 512~\mathrm{TeV}$) reactions, and a specifically built MC model for diffractive processes based on \cite{1983PhR...101..169G,1995PhLB..358..379G,1999PhRvD..59k4017G}. The resonant production of pions (through the resonances $\Delta(1232)$ and res(1600)) is also taken into account in this model, as described in detail in \cite{Kamae:2006bf}.
We include in~\dragon, as an external library, the public code available at \href{https://github.com/niklask/cparamlib}{this link}, which provides the inclusive differential cross sections for positron and electron production in pp reactions. 

Within this model, we parametrize the reactions involving Helium (either as the projectile or as the target) in terms of an empirical rescaling of the cross section relative to the proton-proton channel. In particular, we adopt the rescaling proposed in \cite{2007NIMPB.254..187N}, where the cross sections for the production of $\pi^+$ and $\pi^-$, in a collision between a projectile (with mass number $A_P$) and a target (with mass number $A_T$) are given by:
\begin{equation}
\begin{aligned}
&\sigma_{AA\rightarrow \pi^+X} = (A_P A_T)^{2.2/3}\sigma_{pp \rightarrow \pi^+X}\\
& \sigma_{AA\rightarrow \pi^-X} = \frac{\sigma_{AA\rightarrow \pi^+X}}{K}
\end{aligned}
\end{equation}
with:
\begin{equation}
K = \frac{Z_PZ_T}{A_PA_T} \frac{\sigma_{pp\rightarrow \pi^+X}}{\sigma_{pp\rightarrow \pi^-X}} + \frac{(A_P - Z_p)(A_T-Z_T)}{A_PA_T} \frac{\sigma_{pp\rightarrow \pi^-X}}{\sigma_{pp\rightarrow \pi^+X}}
\end{equation}
and we further assume that the cross section for the production of electrons (positrons) is characterised by the same scaling as the cross section for the production of $\pi^-$ ($\pi^+$). 

\item[Huang2006] This model~\cite{Huang:2006bp} is based on the Monte Carlo event generator DPMJET-III, that is used in the regime $E_p \ge 20$ GeV (at lower energies, where DPMJET-III appears to be not reliable, the same analytical parameterizations used in the Kamae2006 model are adopted).
Also the reactions that involve helium are modeled by means of the same Monte Carlo event generator. 
The differential cross sections for the production of positrons and electrons in the different processes are publicly available at \href{http://www.app.physik.uni-potsdam.de/lepton-prod.html}{this link}.

\end{description}

\begin{figure}
\begin{center}
\includegraphics[width=0.48\columnwidth]{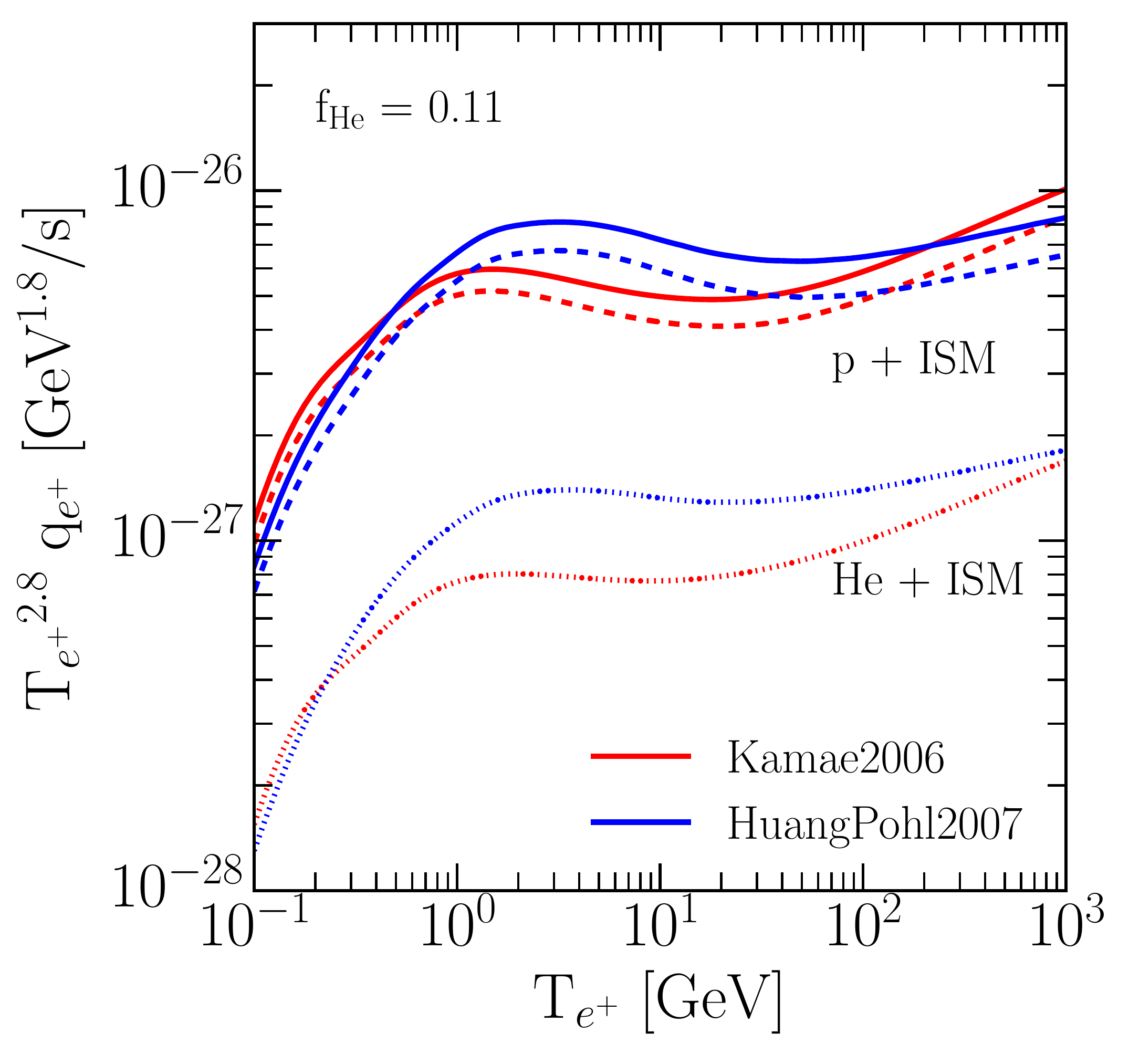} 
\hspace{\stretch{1}}
\includegraphics[width=0.48\columnwidth]{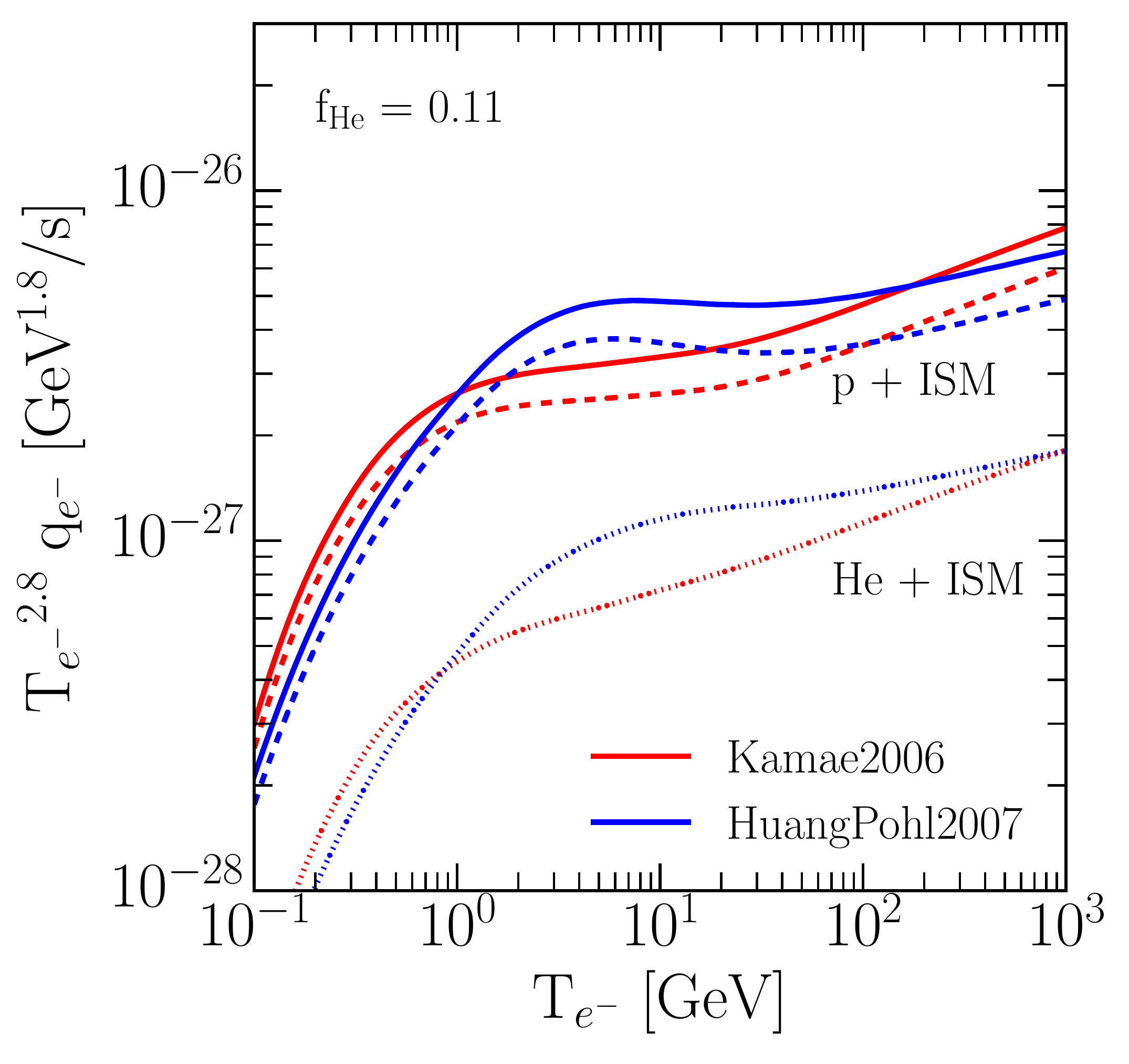} 
\end{center}
\caption{Secondary positron production term (left), electron (right). The solid lines include the contribution of helium both in the targets and in the primary flux.}
\label{fig:lepton_sorce}
\end{figure}

The source terms per unit of hydrogen atom for electrons and positrons, as given by the Kamae and Huang models, are plotted in figure~\ref{fig:lepton_sorce}. The source terms have been obtained by assuming the proton and the helium LIS determined by~\cite{2016A&A...591A..94G}.

Together with the mechanism of their production, it is important to model also the mechanism through which positrons can annihilate as they propagate across the ISM. Such a process can take place as positrons scatter off free electrons:
\begin{equation}
e^+ + e^- \rightarrow \gamma \gamma.
\end{equation} 

The cross section associated to this reaction has been computed by Dirac in \cite{1930PCPS...26..361D} and can be written as:
\begin{equation}
\sigma_{\mathrm{ann}} = \frac{\pi r_e^2}{\gamma + 1} \left[ \frac{\gamma^2 + 4\gamma + 1}{\gamma^2-1} \mathrm{ln} \left( \gamma + \sqrt{\gamma^2 -1}\right) - \frac{\gamma+3}{\sqrt{\gamma^2-1}}\right]
\end{equation}
where $\gamma$ is the positron Lorentz factor, while $r_e$ is the classical electron radius.

The cross section described above enters in the positron transport equation as a ``fragmentation'' term:
\begin{equation}
\frac{1}{\tau^{\rm ann}} = \sigma_{\mathrm{ann}} c \, n_e
\end{equation}
where $n_e$ is the local free electron density.
As it has been found in \cite{Gaggero:1412852}, including the annihilation process in the transport of positrons can impact the very low energy tail (i.e., around 10 MeV) of the interstellar positron flux up to 10\%.

%% file: antiprotons.tex
\section{Secondary antiprotons}\label{sec:antiprotons}

Similarly to what discussed in the lepton section, in the antiproton case as well one can rely on two different approaches: semi-empirical parametrizations tuned on experimental data and Monte Carlo event generators.  

Concerning the former method, the parameterisation that was most in use until recently is the one obtained by Tan and Ng~\cite{Tan:1982nc} which is tuned on a series of measurements performed mainly in the 1960s and 70s. 
More recently, in \cite{Duperray:2003bd,diMauro:2014zea,Kappl:2014hha,Winkler:2017xor}, new parametrizations have been derived from the data provided by the NA49~\cite{Fischer:2003xh} and BRAHMS~\cite{Arsene:2007jd} experiments, which performed accurate measurements of the antiproton production in $p$$p$ collisions, in the antiproton energy range from $4$ to $550$ GeV.  
In this energy range, the accuracy of the analytical parametrizations in reproducing the data has been shown to be remarkably good (e.g., about 10\% in the case of \cite{diMauro:2014zea}). However, when extrapolated beyond the energy range of the available data, the different parameterizations differ of at least $50$\%~\cite{diMauro:2014zea}.

On the other hand, as far as the use of Monte Carlo event generators is concerned, several codes (EPOS LHC, EPOS 1.99 \cite{PhysRevC.92.034906}, SIBYLL\cite{Engel:1999db}, and QGSJET-II-04\cite{PhysRevD.83.014018}) have been used to simulate both $pp$ and $p$He processes and compute the relevant cross sections. These MC generators are widely used in the simulation of extensive CR air showers and have been recently tuned to reproduce minimum bias LHC Run-1 data~\cite{PhysRevD.83.014018,PhysRevC.92.034906}. Some additional tuning is however necessary to reproduce the recent $\bar{p}$ data reported by NA49, BRAHMS and ALICE~\cite{2016PhRvD..94l3007F}.

When computing the antiproton production in the different processes under consideration, it is important to properly take into account the contribution that comes from the decay of secondary antineutrons.
In the past (e.g.,~\cite{Tan:1982nc,Duperray:2003bd}) this was typically done by multiplying the antiproton yield by a factor of 2. 
An approach of this kind is obviously neglecting any possible isospin effect that could make the two cross sections different. Such an effect has been observed by the NA49 collaboration~\cite{Fischer:2003xh} which has  reported an isospin-dependence in the measurements of secondary yields in $n-p$ and $p-p$ collisions. 
This results in $\sigma_{pp \rightarrow \bar{n}} = k_n \sigma_{pp \rightarrow \bar{p}}$ with $k_n \approx 1.5$ for $x_F \approx 0$\footnote{$x_F$ is a Feynman scaling variable and it is defined as $x_F=2p^{*}_L/\sqrt{s}$ where $p^{*}_L$ is the antiproton longitudinal momentum}, although the effect depends on $x_F$ to some extent.
Given the still rudimentary knowledge of these effects, an energy-independent rescaling has been considered by different authors, e.g., $k \approx 1.3 \pm 0.2$ in~\cite{diMauro:2014zea} or $k \approx 1.2 \pm 0.2$ in~\cite{Kappl:2014hha}. 
The authors of~\cite{2016PhRvD..94l3007F}, by making use of EPOS-LHC MC simulations, found that the ratio $k$ is not constant over the phase space and it ranges from 1 to $\sim$1.9. 

As for the case of leptons, also in the case of secondary antiprotons, an important contribution comes from the reactions that involve helium either as a primary CR species or as a target in the ISM.
Until very recently, the cross sections associated to these processes have been plagued by large uncertainties. In fact, because of the absence of experimental data on $p$He collisions, predictions on the cross section associated to this process were made by interpolating between $p$-$p$ and heavier nuclei cross sections like p-C, Cu, Al, Pb and Be (see \cite{Duperray:2003bd} for a collection of these datasets). This situation is expected to see a significant improvement now that the LCHb experiment has started a systematic investigation of the antimatter production in $p$He collisions \cite{LHCb:2017tqz}. 

With all of this considered, in \dragon~three different models to describe $\bar{p}$ production in spallation processes:  

\begin{description}

\item[DiMauro2014]

The differential cross section can be obtained from:
\begin{equation}\label{Eq:intpseudo}
\frac{d\sigma}{dT_{\bar{p}}} = 2 \pi p_{\bar{p}} \int_0^\infty \! d\eta \, \frac{1}{\cosh^2 (\eta)} \sigma_{\rm inv}
\end{equation}
where $\eta=-\ln \left[ \tan(\theta/2) \right]$ is the pseudo-rapidity defined in terms of the the scattering angle ($\theta$) and the invariant cross section:
\begin{equation}
\sigma_{\rm inv} \equiv E \frac{d^3\sigma}{dp^3} (\sqrt{s}, x_{\rm R}, p_{\rm T})
\end{equation}
depends only on Lorentz invariant quantities: the center of mass (CM) energy ($\sqrt{s}$), the transverse momentum of the produced antiproton ($p_{\rm T}$) and the ratio of the $\bar{p}$ energy to the maximally possible energy in the CM frame ($x_{\rm R}$).
The lower limit in the integral in~\ref{Eq:intpseudo} is set to $0$ since it corresponds, in good approximation, to the kinetic limits of the angular integration~\cite{2017PhRvD..96d3007D}.

In~\cite{diMauro:2014zea}, an analytic formula is proposed to reproduce the new NA49 data and with an explicit dependence on $s$:
\begin{eqnarray}
\sigma_{\rm inv} [{\rm mb}] & = & \sigma_{\rm in}(s) (1-x_R)^{C_1} e^{-C_2 x_R} \nonumber \\
& \times & || C_3 (\sqrt{s})^{C_4} e^{-C_5 p_T} +  \cdot C_6 (\sqrt{s})^{C_7} e^{-C_8 p_T^2} + C_9 (\sqrt{s})^{C_{10}} e^{-C_{11} p_T^3} || 
\label{eq:DiMauro}
\end{eqnarray}
where $\sigma_{\rm in}(s)$ is the inelastic proton cross section and is defined as the difference between the total $pp$ scattering cross section and its elastic counterpart.
In~\cite{diMauro:2014zea}, $\mathrm{\sigma_{in}^{pp}}$ is obtained by fitting the latest data provided by the Particle Data Group (PDG) on the total and elastic $pp$ cross sections~\citep{PDGcs2013}.

To describe proton-nucleus production within this model we adopt the formula given in~\cite{Duperray:2003bd} which reproduces the experimental cross sections to within a few tens of percent for incident energies from 12 GeV up to 400 GeV, and for target mass $1 \le A \le 208$:
\begin{equation}
\label{eq:heliumap}
\sigma_{\rm inv} [{\rm mb}] = \sigma_{\rm in} A^{C_1 \ln (\sqrt{s} / C_2) p_{\rm T}} (1-x_{\rm R})^{C_3 \ln(\sqrt{s})} {\rm e}^{-C_4 x_{\rm R}} \left[ C_5 \sqrt{s}^{C_6} {\rm e}^{-C_7 p_{\rm T}} + C_8 \sqrt{s}^{C_9} {\rm e}^{C_{10} p_{\rm T}^2} \right]
\end{equation}
where $A$ is the target mass and $\sigma_{\rm in}$ is the total inelastic cross section for pA collisions.

\item[Winkler2017]

In \cite{Winkler:2017xor}, the author has evaluated the antiproton production cross sections by fitting several analytical parametrisations to proton-proton scattering data of RHIC and LHC. 
The violation of Feynman scaling as well as an enhanced strange hyperon production (suggested by experimental data on $\Lambda^-$ production) are investigated and are found to increase the antiproton cross sections at high energies with respect to similar analysis.

Finally, he employed proton-proton, neutron-proton and proton-nucleus scattering data to determine the strength of isospin effects which induce an asymmetry between antiproton and antineutron production. 

In~\dragon, we interpolate the publicly available tables of cross sections for p-p, p-He,  He-p and He-He scattering\footnote{\url{https://arxiv.org/abs/1701.04866}}.

\item[Feng2016] In this model, Monte-Carlo generators (such as EPOS LHC, EPOS 1.99, SIBYLL, and QGSJET-II-04) and accelerator data (NA49, BRAHMS, and ALICE) are used to assess the antiproton production cross sections and their uncertainties.
The comparison of data with MC generators presented in~\cite{2016PhRvD..94l3007F} shows a better agreement with EPOS LHC and QGSJET-II-04 with respect to SIBYLL at all $p_{\rm T}$-values. In~\dragon~we then implement only these two cases.
The production cross sections for the channel $pp$, $p$He, He$p$, and HeHe have been provided by the authors in a ROOT file.

\end{description}

We remark here that while in \dragon~we adopt~\cite{2016PhRvD..94l3007F} as a reference for the antiproton production within MC event generators, other relevant results have been obtained in the recent literature. In particular, we mention the analysis carried out in~\cite{2015ApJ...803...54K}, where a modification of the QGSJET-II-04 hadronization model (denoted as QGSJET-II-04m) has been introduced in order to make QGSJET suitable to treat collisions occurring in the low-energy regime.

\begin{figure}
\begin{center}
\includegraphics[width = 0.48 \columnwidth]{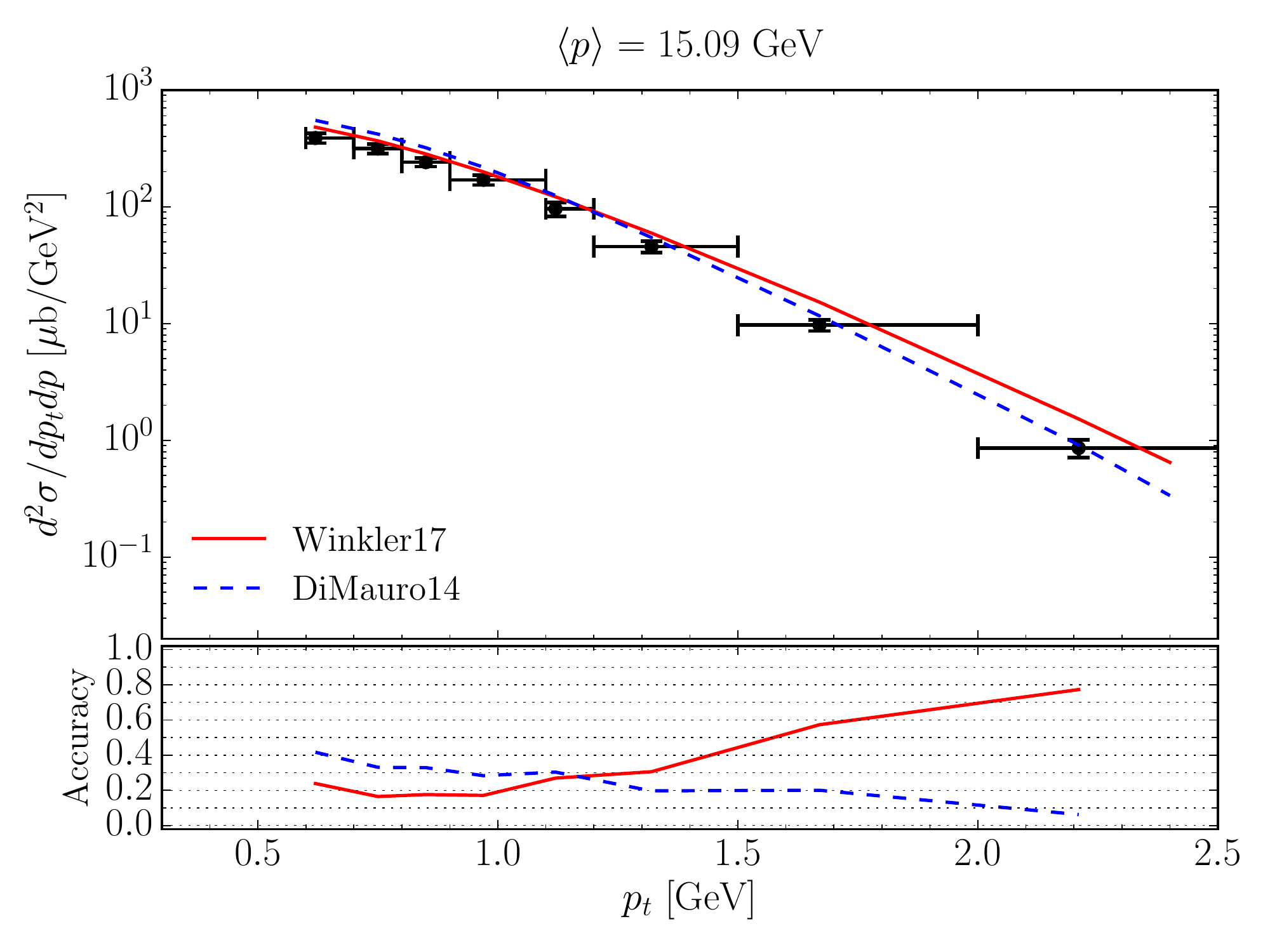}
\includegraphics[width = 0.48 \columnwidth]{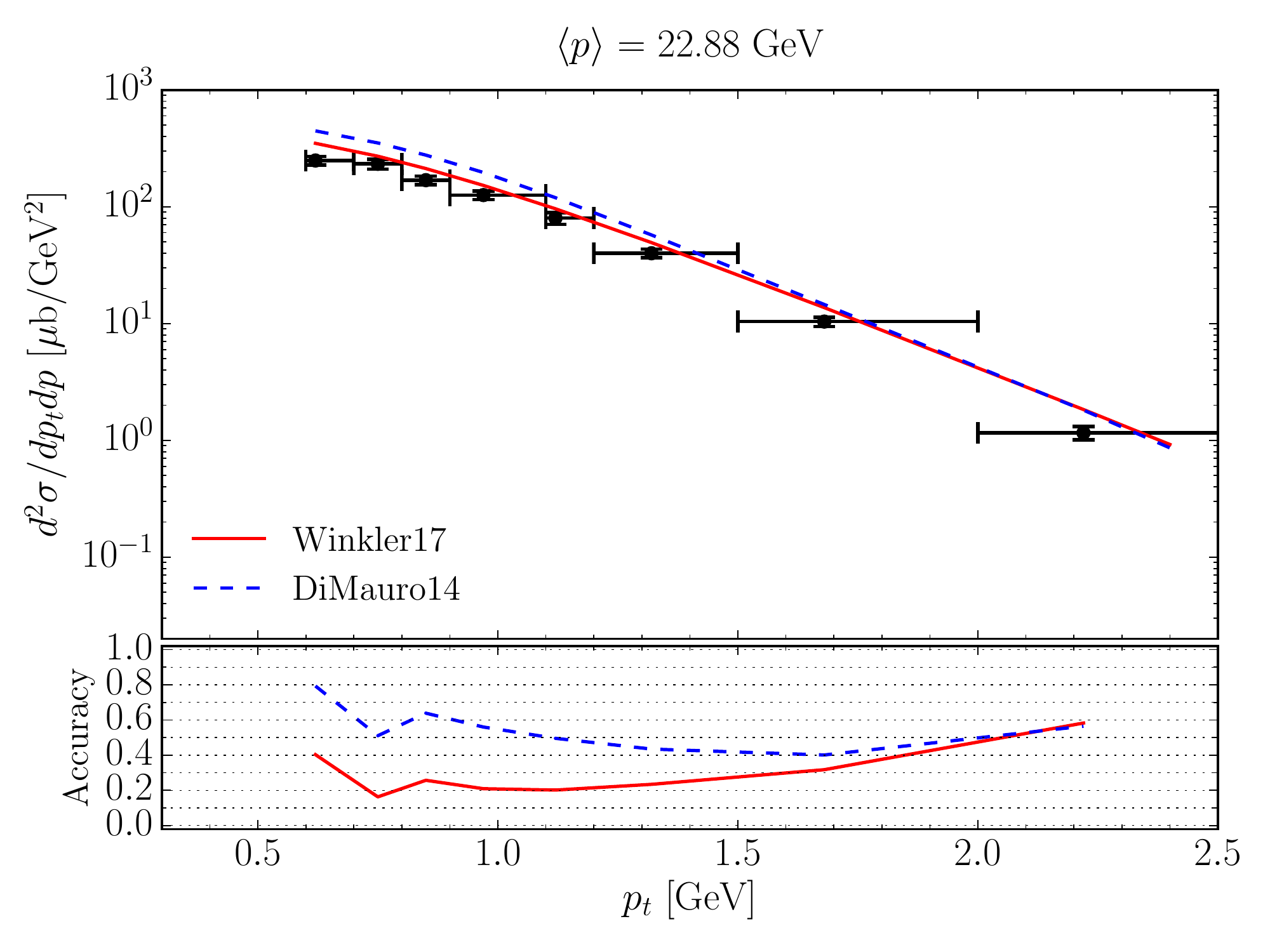}\\
\includegraphics[width = 0.48 \columnwidth]{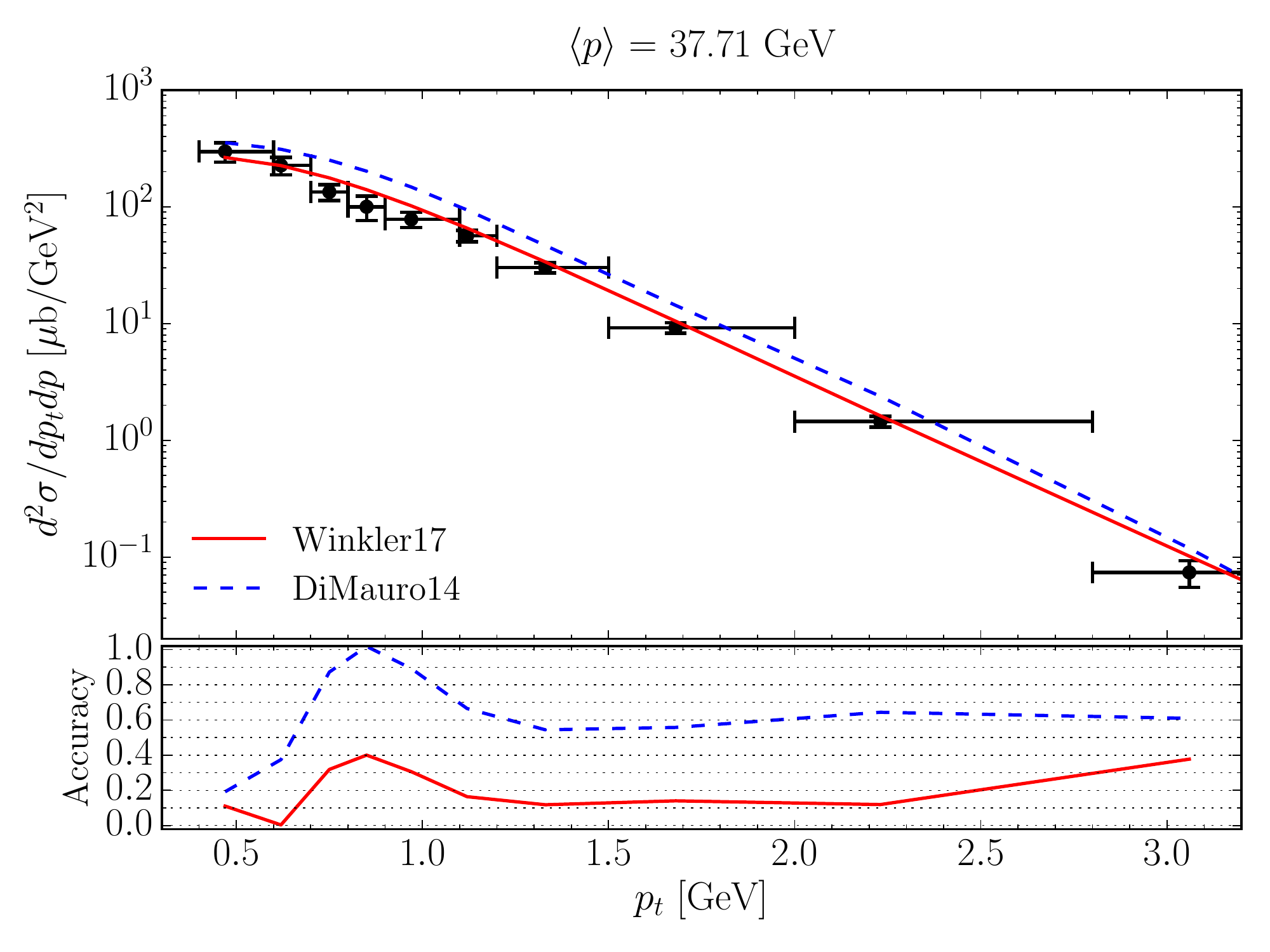}
\includegraphics[width = 0.48 \columnwidth]{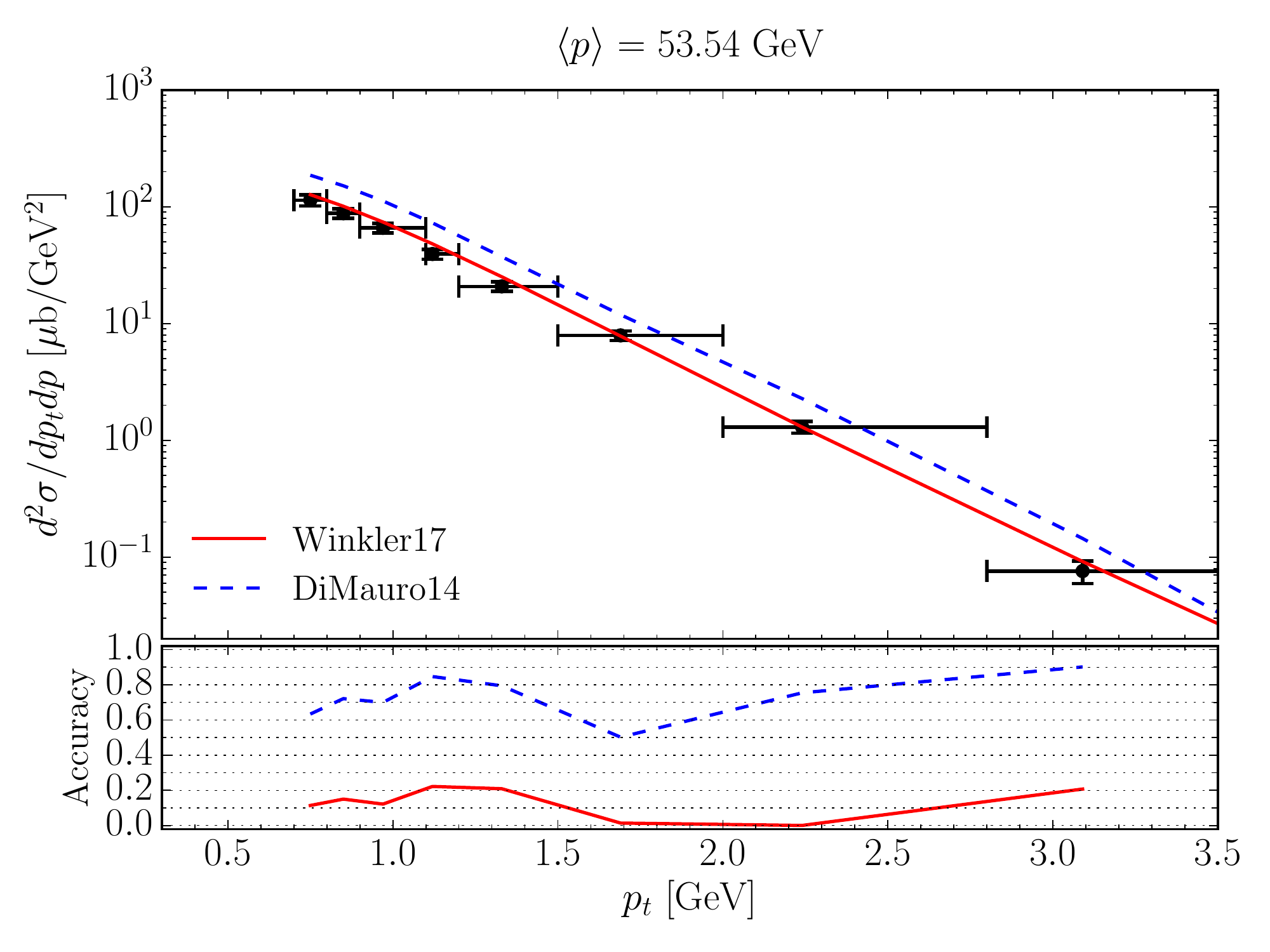}
\includegraphics[width = 0.48 \columnwidth]{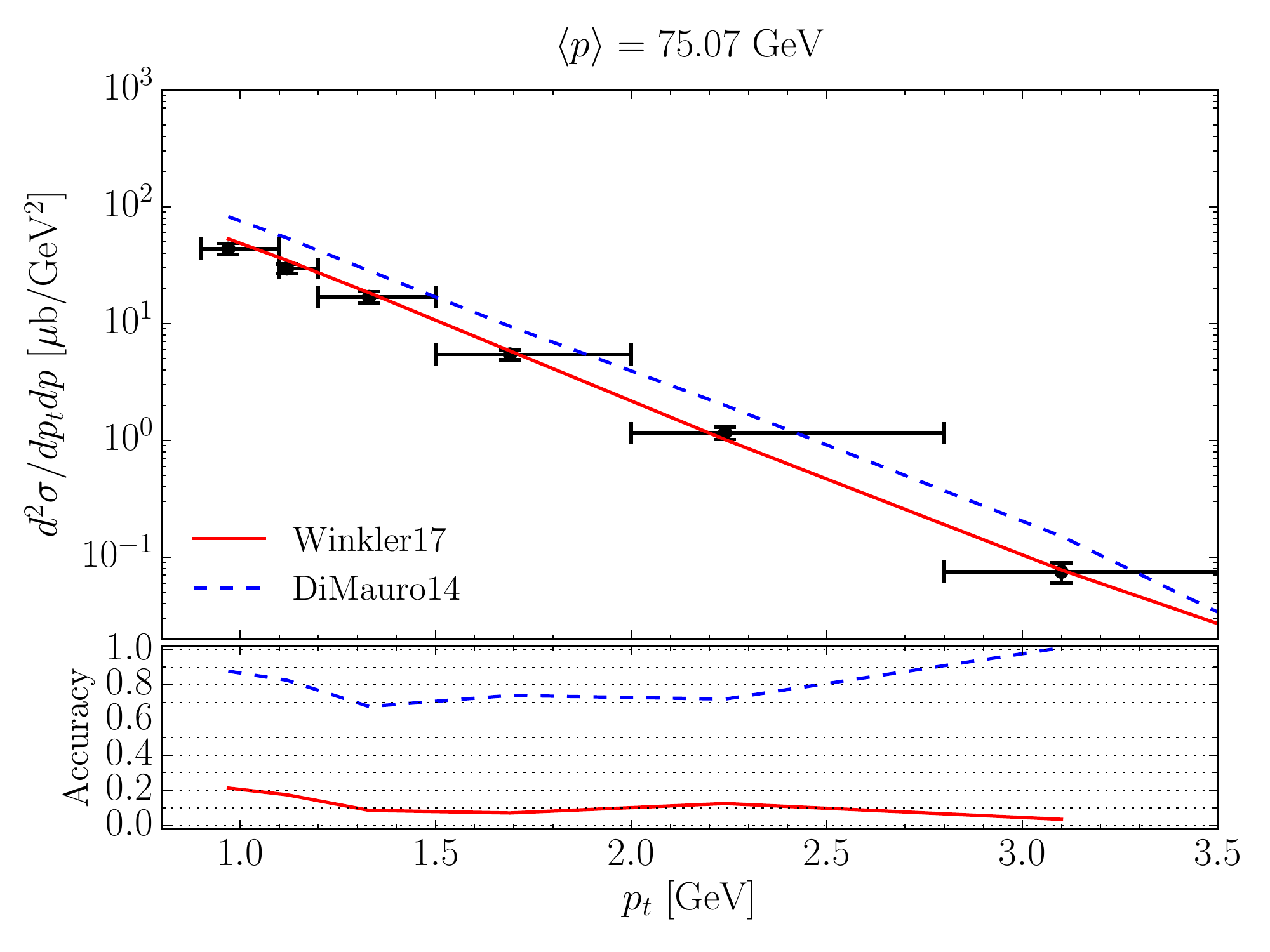}
\includegraphics[width = 0.48 \columnwidth]{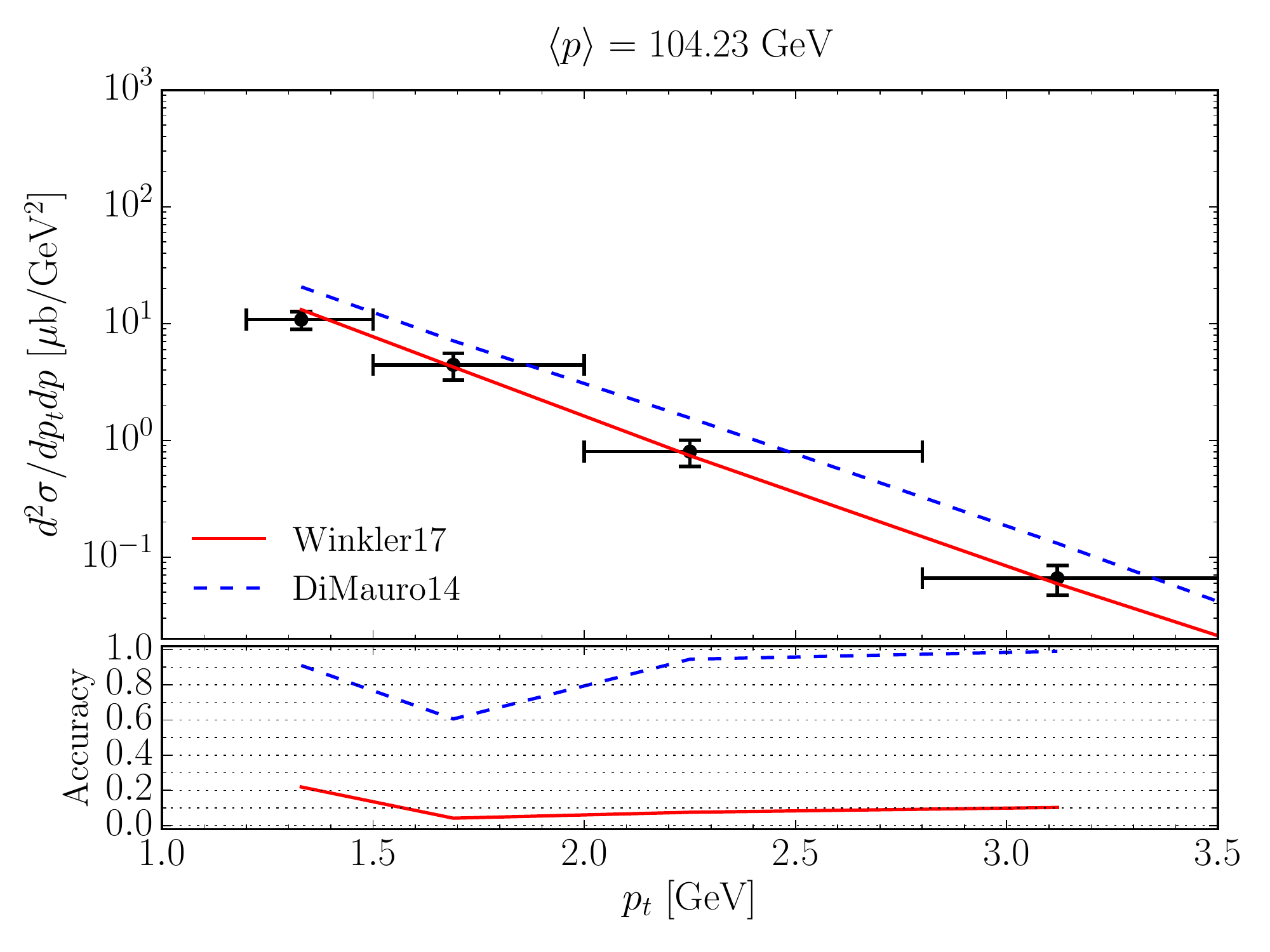}
\end{center}
\caption{Double differential antiproton production cross section in the pHe channel at $\sqrt{s}$ = 110 GeV as a function of the antiproton transverse momentum $p_t$, as measured by the LHCb experiment (black points) compared with the predictions of the {\tt Winkler2017} (red solid lines) and {\tt DiMauro2014} (blue dashed lines) models. The different plots correspond to the different bins in the antiproton total momentum (for each bin we report the average momentum $\langle p \rangle$) in the label above the plot). The panels below the plots show the accuracy of the two models in reproducing data, intended as $\left| \mathrm{data} - \mathrm{model}\right|$/data. \vspace{1cm}}
\label{fig:pHe_LHCb}
\end{figure}

\begin{figure}
\begin{center}
\includegraphics[width=0.49\columnwidth]{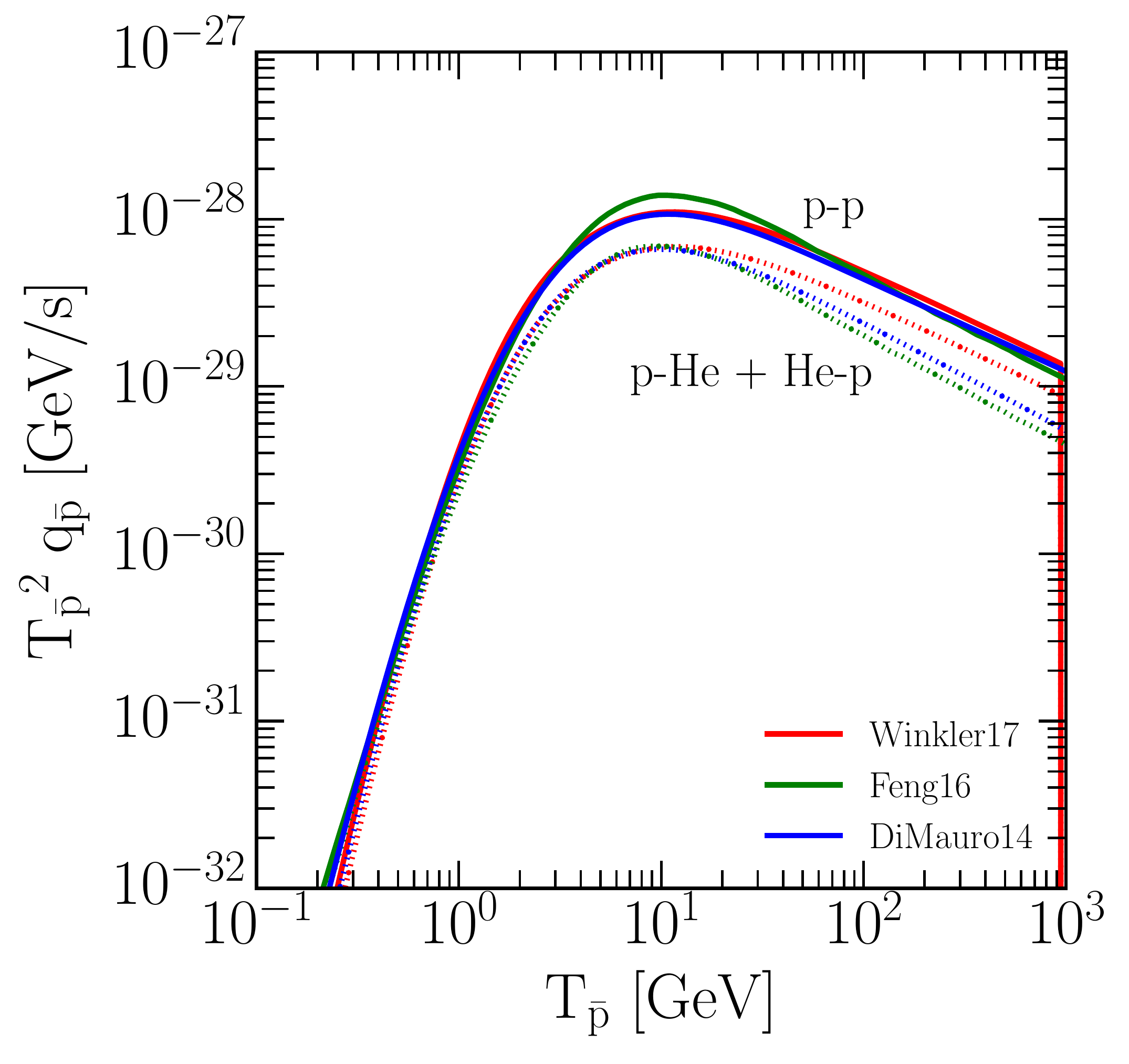} 
\includegraphics[width=0.49\columnwidth]{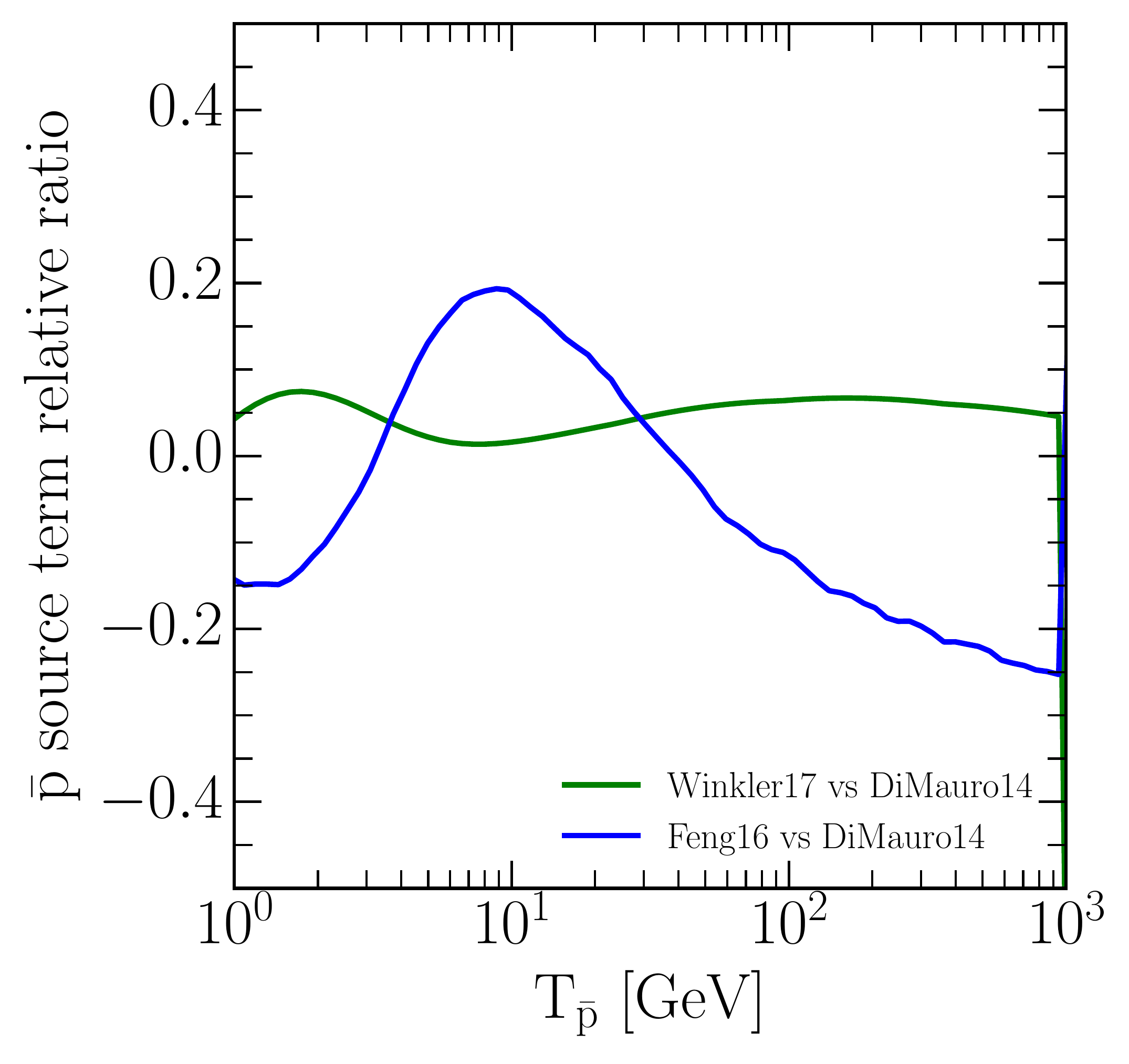} 
\end{center}
\caption{Left panel: the antiproton source term for the cross-section models in~\dragon. For Feng+16 we plot only the EPOS-LHC model. The solid lines show the source term due to pp interactions, the dotted lines the total contribution from He (in the target and in CRs). Right panel: the relative ratio of the antiproton source term for the models {\tt Winkler2017} and {\tt Feng2016} models with respect to {\tt DiMauro2014}.}
\label{fig:figcompsourceterm}
\end{figure}

As mentioned above, the LHCb experiment has recently released the results of the first measurement of the antiproton production cross section in $p$He collisions \cite{LHCb:2017tqz} at $\sqrt{s}$ = 110 GeV. We use these data to investigate the accuracy of the {\tt Winkler2017} and {\tt DiMauro2014} parametrisations\footnote{We leave aside here the {\tt Feng2016} model, as a detailed comparison between LHCb data and the results of the MC event generators included in this model (EPOS LHC, EPOS 1.99, QGSJET-II-04) has already been carried out in \cite{LHCb:2017tqz}.}. Results are shown in figure~\ref{fig:pHe_LHCb}, where the predictions given by the two parametrisations are compared with a representative set of the LHCb results. As it can be seen, once that the uncertainty in the determination of the antiproton transverse momentum $p_t$ is taken into account, both models are in a relatively good agreement with LHCb data. It appears also that, while at low antiproton momenta (approximately $\langle p\rangle<$ 30 GeV) the performances of the two models can be considered to be comparable (the relative displacement between predictions and data ranging between the 20\% and the 80\%), at higher energies the {\tt Winkler2017} model is in a better agreement with data, with an accuracy that in vast regions of the ($\langle p\rangle$,$p_t$) parameter space is even around, or better than, the 10\% level. A more detailed discussion about the impact of LHCb data on the parametrisations of the antiproton production cross section will be presented in~\cite{Korsmeier2017bis}.   

In figure~\ref{fig:figcompsourceterm} the contributions to the total antiproton source term coming from the $pp$, He$p$ and $p$He channels are shown for the different models. As it can be seen, the difference between the approaches can amount up to 20\% at around 10~GeV.

As the other CR species, also antiprotons can interact with the particles of the ISM as they propagate through the Galaxy. As a result of such interactions, antiprotons can either annihilate or simply lose a fraction of their energy~\cite{2002ApJ...565..280M}.

The former process is described by the term:
\begin{equation}
\frac{1}{\tau^{\rm f}_{\bar{p}}(T)} = 
\beta(T) c n_{\rm H} \left[ \sigma_{\rm H,\bar{p}}(T) + f_{\rm He} \sigma_{\rm He,\bar{p}}(T) \right] 
\end{equation}
where $\sigma_{\mathrm{k},\bar{p}}$ is the total inelastic cross section associated to the $\bar{p}$-k collision.
In the case of hydrogen, this cross section is given as a function of the antiproton rigidity ($R$ in GV) by~\cite{2008ApJ...678..907S}:
\begin{equation}
\sigma_{\rm H,\bar{p}} [\text{mb}] = -107.9 + 29.43 \ln R -1.655 \ln^2 R + 189.9 R^{-1/3}.
\end{equation}
In the case of helium, we adopt the prescription proposed in~\cite{1997APh.....6..379M}, where the inelastic scattering of $\bar{p}$ on target nuclei of mass $A$ is given by: 
\begin{multline}
\sigma_{\rm He,\bar{p}} [\text{mb}] = A^{2/3} \times \left[ 48.2 + 19 (T - 0.02)^{-0.55} \right. \\ \left. - 0.106 \, A^{0.927} \, T^{-1.2} + 0.059 \, A^{0.927} + 0.00042 \, A^{1.854} \, T^{-1.5} \right]
\end{multline}

Non-annihilation inelastic interactions of antiprotons with interstellar protons yield lower energy antiprotons in the final state. 
This component dominates the local flux at low energies.
As in~\cite{1999ApJ...526..215B}, we treat inelastically scattered secondary antiprotons as a separate ``tertiary'' component whose source function is in the form:
\begin{equation}
\Gamma^{}_{\bar{p}^{\rm s} \rightarrow \bar{p}^{\rm t}}(T) =
c \, n_{\rm H} \int \! dT' \, \beta(T') N_{\bar{p}^{\rm s}}(T') 
\left[ \frac{d\sigma_{\rm H,\bar{p}^{\rm s} \rightarrow \bar{p}^{\rm t}}}{dT}(T,T') 
+ f_{\rm He} \frac{d\sigma_{\rm He,\bar{p}^{\rm s} \rightarrow \bar{p}^{\rm t}}}{dT}(T,T')  \right]
\end{equation}
notice that the tertiary source term depends on the secondary antiproton density which we obtain beforehand in the nuclear chain. 

\begin{figure}[t]
\begin{center}
\includegraphics[width=0.48\columnwidth]{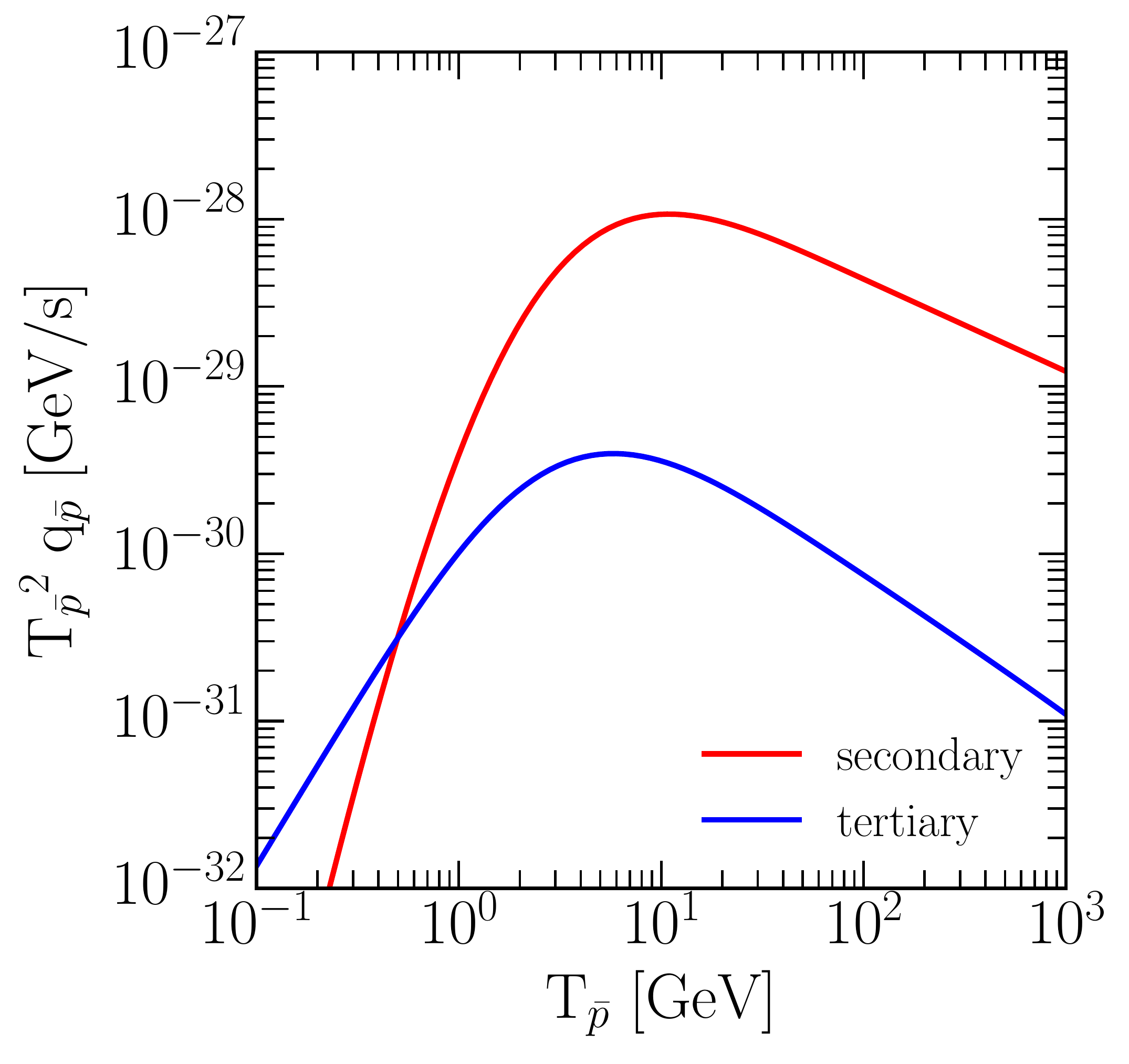} 
\end{center}
\caption{Secondary and tertiary antiproton source term computed according the {\tt DiMauro2014} model. We assume the proton and helium LIS derived in~\cite{2016A&A...591A..94G}.}
\label{fig:secterap}
\end{figure}

The antiproton density after propagation is given by the sum of the secondary and tertiary components. 
This procedure provides an accurate solution as long as the transport equation is linear.
The differential cross section is taken in agreement with~\cite{Tan:1982nc}:
\begin{equation}
\frac{d\sigma_{\rm H,\bar{p}^{\rm s} \rightarrow \bar{p}^{\rm t}}}{dT}(T,T') = \frac{\sigma_{\rm non-ann}(T')}{T'}
\end{equation}
where $\sigma_{\rm non-ann}$ is obtained as the difference between the total inelastic cross section and the inelastic annihilation cross section:
\begin{equation*}
\sigma_{p\bar{p}} [\text{mb}] = 
\begin{cases}
661 \left( 1 + 0.0115 \, T^{-0.774} - 0.948 \, T^{0.0151} \right) & T < 15.5~\text{GeV} \\
36 \, T^{-0.5} & T \ge 15.5~\text{GeV}
\end{cases}
\end{equation*}
and we rescale with $A^{2/3}$ for taking into account helium in the target.

We stress that tertiary antiprotons can in turn be subject to fragmentation, however further terms in the series, such as antiprotons produced by tertiary interactions constitutes a subdominant component. 

A comparison between secondary and tertiary antiproton source terms is shown in figure~\ref{fig:secterap}.

\begin{center}
\begin{table}[t]
\scriptsize
\begin{tabular}{ccccccccccc}
\toprule
C$_1$ & C$_2$ & C$_3$ & C$_4$ & C$_5$ & C$_6$ & C$_7$ & C$_8$ & C$_9$ & C$_{10}$ & C$_{11}$ \\ 
\midrule 
$4.448$ & $3.735$ & $5.02$e$-3$ & $0.708$ & $3.527$ & $0.236$ & $-0.729$ & $2.517$ & $-1.822$e$-11$ & $3.527$ & $0.384$ \\ 
$0.1699$ & $10.28$ & $2.269$ & $3.707$ & $9.205$e$^{-3}$ & $0.4812$ & $3.360$ & $0.06394$ & $-0.1824$ & $2.485$ & \\
\bottomrule
\end{tabular}
\label{table:1}
\caption{Best-fit parameters of equation~\ref{eq:DiMauro} and~\ref{eq:heliumap}.}
\end{table}
\end{center}

%% file: conclusions.tex
\section{Conclusions}

In the second paper of a series (see also~\cite{2017JCAP...02..015E}) we presented a detailed and comprehensive description of how we implement the network of nuclear interactions between charged Galactic CRs and the interstellar gas in \dragon.

We first discussed and compared two widely used parametrizations for the total inelastic scattering, which are in good agreement with each other, and then presented a procedure to determine the fragmentation cross sections in the energy range 100 MeV/n to 100 GeV/n, mainly based on a set of parameterizations produced by Webber and colleagues (properly re-tuned to available data).

We then focused our attention to the production of antiparticles, given their crucial role in CR physics, e.g., as background for dark matter searches.

For positron (and secondary electron) production we compared the results obtained by~\cite{Kamae:2006bf} with those in \cite{Huang:2006bp}.
The differences between the two models computed by means of different Monte Carlo codes are still sizable being up to 50\% in the 1-100~GeV energy range.
Regarding antiprotons, we first used the results of~\cite{2016PhRvD..94l3007F} to compute the $\bar{p}$ source term by different Monte Carlo generators. 
We then considered the models presented in~\cite{diMauro:2014zea} and \cite{Winkler:2017xor}, who provided their production cross sections in terms of a global fit of $\bar{p}$ production measurements. 
We obtained differences among the models of $\sim 20$\%. Furthermore, for the first time, we employed the very recent LHCb data to investigate the validity of the two parametrisation in describing antiproton production in $p$He reactions. 
 
Current experiments are measuring the high-energy charged CR spectra with unprecedented accuracy. However, the inclusive and partial cross sections of CRs with the interstellar gas still induce a significant uncertainty on the interpretation of these great data, thus weakening our ability to reconstruct the CR parentage and origin.  
An extended program of cross section measurements at energies much larger than GeV/n, either for hydrogen and helium targets, and interpreted by modern Monte Carlo codes, as FLUKA~\cite{2014NDS...120..211B} or GEANT4~\cite{2003NIMPA.506..250A}, appears now compelling in order to provide a much more robust estimate of the uncertainties associated to these processes affecting our understanding of CR transport processes.

%% file: acknowledgments.tex
\newpage \section*{Acknowledgments}

We acknowledge:
\begin{itemize}
\item the use of the {\tt CRDB}~\cite{2014A&A...569A..32M} database for CR measurements. 

\item the use of the following codes: {\tt CROSEC} by Barashenkov and Polanski, {\tt WNEWTR} by Webber and {\tt YIELDX} by Silberberg and Tsao.

\item the use of the table of cross section measurements in \url{isotope_cs.dat} from {\tt GALPROPv54} at \url{galprop.stanford.edu}. 

\item the {\tt cparamlib} library for calculation of fluxes for stable secondary particles created in proton-proton interactions.

\end{itemize}

We thank Prof.~Webber for providing us the fragmentation cross-section tables from an updated version of his code we use in \S~\ref{sec:frag}.

We are indebted to 
Fiorenza Donato, 
David Maurin, 
Pierre Salati,
Nicola Tomassetti, 
Piero Ullio for numerous discussions and suggestions.

CE acknowledges the European Commission for support under the H2020-MSCA-IF-2016 action, grant No. 751311 GRAPES – Galactic cosmic RAy Propagation: an Extensive Study.

AV acknowledges support from the German-Israeli Foundation for Scientific Research.

MDM acknowledges support by the NASA Fermi Guest Investigator Program 2014 through the Fermi multi-year Large Program N. 81303 (P.I.~E.~Charles).